\documentclass{aa}  
\usepackage{natbib}
\usepackage{tabularx}
\usepackage{xcolor} 
\usepackage{pdflscape}
\usepackage{float}
\usepackage{placeins}
\usepackage{txfonts}
\usepackage[switch]{lineno}

\newcommand{\fotwosurf}{$f\mathrm{O}_{2}^{\mathrm{surf}}$\xspace}
\newcommand{\fotwomelt}{$f\mathrm{O}_{2}^{\mathrm{melt}}$\xspace}
\newcommand{\fotwo}{$f\mathrm{O}_{2}$\xspace}
\definecolor{mypink}{RGB}{255,105,180} % hot pink

\usepackage[colorlinks=true, citecolor=blue, linkcolor=black, urlcolor=black]{hyperref}
\usepackage{graphicx}
\usepackage{txfonts}

\begin{document} 
\linenumbers
\setlength\linenumbersep{10pt}
\title{Geophysical and atmospheric implications of $f$O$_{2}$-dependent melting on rocky exoplanets}

   \author{Mariana Sastre \inst{1}\corrauth{m.c.villamil.sastre@rug.nl}, Tim Lichtenberg \inst{1}, Laurent Soucasse\inst{2,3}, Dan J. Bower \inst{4}, Harrison Nicholls\inst{5}, Inga Kamp\inst{1}}

   \institute{Kapteyn Astronomical Institute,  University of Groningen, Groningen, The Netherlands
              \and
              Netherlands eScience Center, Amsterdam, The Netherlands
              \and
              Interuniversity Micro Electronics Centre, Leuven, Belgium
              \and
              Institute of Geochemistry and Petrology, Department of Earth and Planetary Sciences, ETH Z\"urich, Z\"urich, Switzerland
              \and
              Institute of Astronomy, University of Cambridge, Cambridge, United Kingdom
             }

   \date{}

\authorrunning{Sastre et al.}

\abstract{
The geochemical evolution of long-lived magma oceans is strongly regulated by volatile exchange between the molten mantle and the atmosphere. For planets inside the runaway-greenhouse limit, this coupled evolution can persist for billions of years, governing bulk density, surface conditions, and long-term geodynamics. However, most existing studies assume Earth-like (oxidized) conditions and neglect the influence of redox state on melt thermodynamics and volatile release. We quantified how experimentally derived, oxygen-fugacity-dependent melting curves implemented within the coupled interior-atmosphere framework \texttt{PROTEUS} propagate into the thermal structure, melt fraction, and rheological evolution of rocky exoplanet interiors, applying this to the short-period super-Earth GJ~1132~b. We found strongly non-linear thermal responses to variations in melting curves. In volatile-poor systems, reduced melting curves ($f$O$_2$ $\leq$ IW, where IW denotes the iron--wüstite buffer) promote earlier deep-mantle crystallisation relative to oxidised (IW+2.0) and Earth-like (IW$+$4.0) cases (range IW$-$4.0 to IW$+$4.0), favouring late-stage surface magma oceans sustained by greenhouse warming, while oxidized melting curves maintain higher melt fractions and a vertically extended magma ocean. Reduced mantles produce massive H$_2$-CO-rich atmospheres; oxidized mantles favour thinner H$_2$O-CO$_2$ envelopes. In volatile-rich systems, the interior reaches radiative equilibrium at high melt fractions, sustaining a steady-state global magma ocean in which melting curve variations do not significantly influence solidification timing. This indicates a hierarchical control: volatile inventory and surface oxygen fugacity act as the primary regulators of thermal state, while oxygen-fugacity-dependent melting relations provide a secondary modulation. These contrasting regimes produce distinct atmospheric compositions and formation timescales, offering testable spectral predictions for close-in rocky exoplanets evaluable with forthcoming JWST observations.
}


   \maketitle
   \nolinenumbers

\section{Introduction}
\label{sec:intro}
Planetary formation is thought to proceed through various stages of accretion by impacts of planetesimals and the accretion of pebbles within the protoplanetary disc \citep{johansen_forming_2017,raymond_impact-driven_2020, batygin_formation_2023}. The timescales and initial heat budget at which this process occurs imply a significant amount of energy, enough to induce multiple episodes of large-scale mantle melting. 
This stage is known as the magma ocean phase \citep{elkins-tanton_magma_2012}. As the magma ocean cools and crystallizes, some fraction of the volatile content dissolved within the planet's interior is outgassed to the exterior, contributing to the growth and composition of a secondary atmosphere and initiating long-term climate evolution \citep{elkins-tanton_linked_2008, lebrun_thermal_2013,  hamano_emergence_2013,bower_linking_2019, bower_retention_2022, salvador_magma_2023, Lichtenberg2023ASPC,nicholls_magma_2024, boukare_2025}. The outgassed atmosphere strongly influences the crystallisation of the magma ocean. By producing a greenhouse blanketing effect, it retains heat and can keep surface temperatures above the solidus temperature, which is the threshold for melting under given thermodynamic and compositional conditions. This helps maintain a molten surface and extends the lifetime of the magma ocean. \citep{abe_1985,abe_1988,schaefer_predictions_2016, salvador_relative_2017,lichtenberg_vertically_2021}.  

Previous studies \citep{elkins-tanton_linked_2008, lebrun_thermal_2013} showed that volatile species exhibit contrasting partitioning behaviour during magma ocean evolution. Owing to its low solubility in silicate melts, carbon-bearing species preferentially reside in the atmosphere, being efficiently outgassed from the earliest stages of crystallisation \citep{bower_retention_2022}. In contrast, the higher solubility of H$_2$O causes it to remain largely dissolved in the melt, favouring retention within the interior until the final stages of solidification \citep{Suer2023FrEaS,sossi_solubility_2023,bower_retention_2022, Bower2025ApJ}. Water-rich atmospheres enhance thermal blanketing and suppress radiative cooling, extending the molten phase. Therefore, the initial water inventory plays a key role in setting magma ocean lifetimes. In contrast, atmospheres dominated by carbon-bearing species (e.g., CO$_2$, CO), are generally more transparent to outgoing longwave radiation and therefore less effective at retaining heat \citep{kiehl_1997, salvador_magma_2023}. Stellar evolution can produce a similar effect, as high stellar irradiation may suppress efficient cooling through atmospheric buffering of infrared emission, potentially sustaining long-lived magma oceans. However, recent studies indicate that this behaviour is not governed by a universal critical flux threshold but depends more on atmospheric and interior properties \citep{hamano_emergence_2013, hamano_2015, salvador_relative_2017, Boer2025ApJ}.

Exoplanet surveys provide an opportunity to test models of coupled interior–atmosphere evolution during magma ocean phases \citep{Kreidberg2025PNAS,Lichtenberg2025Sci}. Recent JWST observations have transformed our understanding of rocky ultra-short-period exoplanets, revealing that some retain volatile secondary atmospheres rather than being entirely desiccated. Using JWST/NIRCam and MIRI observations, \citet{hu_secondary_2024} detected a persistent volatile envelope around 55 Cnc e, inconsistent with a bare-rock or silicate-vapour atmosphere and likely dominated by CO or CO$_2$. Similarly, \citet{teske_thick_2025} reported the dayside emission spectrum of TOI-561 b, whose brightness temperatures require a thick atmosphere that redistribute heat away from the planet's dayside. These results provide the first evidence that volatile-rich envelopes can survive on highly irradiated rocky planets. Because these highly irradiated planets exist at thermodynamic conditions that may favour persistent dayside or global magma ocean regimes, these detections open a new avenue for linking observable secondary atmospheres to their geochemical histories. Therefore, interpreting these observations requires a detailed understanding of the coupled evolution of planetary interiors and atmospheres, and of the processes that regulate volatile release and retention.

Although atmospheric processes have received substantial attention, the geochemical and geophysical evolution of the interior is equally important in shaping a planet’s long-term climate and thermodynamic state. A key factor in this evolution is the oxidation state of the silicate mantle, which influences melting behaviour, volatile speciation, and interior–atmosphere exchange \citep{hirschmann_magma_2012, lichtenberg_vertically_2021,gaillard_redox_2022, wordsworth_atmospheres_2022,salvador_magma_2023, nicholls_magma_2024}. The mantle redox state reflects the equilibrium distribution of electrons among multivalent elements, particularly the balance between Fe$^{2+}$ and Fe$^{3+}$ in silicate phases \citep{cottrell_2025}. Controlled by the oxygen chemical potential, this thermodynamic state is commonly quantified by the oxygen fugacity, $f\mathrm{O}_{2}$, which represents the thermodynamic availability of oxygen for redox reactions, even in the absence of free molecular oxygen \citep{wade_2005, kasting_2014, salvador_magma_2023}. High oxygen fugacity corresponds to more oxidizing mantle conditions, whereas low oxygen fugacity favours reduced chemical species.

As the dominant multivalent species in the mantle, iron simultaneously records and controls redox conditions through the balance of its oxidation states \citep{salvador_magma_2023}. The oxygen fugacity of a system is set by equilibrium among coexisting redox-sensitive phases, which buffer the chemical potential of oxygen at specific conditions. These equilibrium reactions define reference oxygen fugacity states that are commonly referred to as redox buffers \citep{frost_2008, hirschmann_magma_2012, cottrell_2025}. One of the most commonly used redox buffer is the iron–wüstite (IW) buffer Eq.~\ref{eq:iron-wustite}., which represents the equilibrium between metallic iron (Fe) and ferrous iron incorporated in wüstite (FeO, iron(II) oxide). The oxygen fugacity relative to this buffer is expressed in log units as $\Delta\log f\mathrm{O}_2(\mathrm{IW}) = \log f\mathrm{O}_2 - \log f\mathrm{O}_2^{\mathrm{IW}}$, where $f\mathrm{O}_2^{\mathrm{IW}}$ is the oxygen fugacity at the IW equilibrium at the same pressure and temperature. 

\begin{equation}
\mathrm{Fe}_{\mathrm{metal}} + \tfrac{1}{2}\,\mathrm{O}_{2}
\rightleftharpoons
\mathrm{FeO}_{\mathrm{w\ddot{u}stite}} .
\label{eq:iron-wustite}
\end{equation}

This buffer is characteristic of moderately reducing conditions in planetary interiors. With higher relative \fotwo corresponding to greater stabilization of oxidized iron species and lower values favouring metallic iron and more reducing conditions \citep{salvador_magma_2023}.

Previous studies have shown that oxygen fugacity and crystallisation regimes jointly regulate the thermal and chemical evolution of planetary magma oceans  \citep{oneil_2002, armstrong_speciation_2015, sossi_2020, lichtenberg_vertically_2021, bower_retention_2022, gaillard_redox_2022, shorttle_distinguishing_2024, nicholls_magma_2024}. Under reducing conditions relative to the IW buffer, coupled interior–atmosphere models predict C- and H-rich atmospheres dominated by H$_2$ and CO, whereas oxidizing conditions favour H$_2$O- and CO$_2$-dominated compositions. CO remains stable across a wide redox range.

Complementary to these redox-driven trends, melt–solid separation modulates evolution \citep{boukare_2025}. For instance, \citet{bower_retention_2022} showed that fractional crystallisation (crystal segregation) and equilibrium crystallisation (no crystal segregation) lead to different melt redistribution, affecting the timing of cooling and outgassing processes.

In addition, redox-dependent changes in atmospheric speciation influence radiative properties and thermal evolution. For example, by treating $f$O$_2$ as a free parameter, \citet{nikolaou_what_2019, katyal_effect_2020} showed that changes in the oxidation state modify the equilibrium between interior and atmosphere, alter volatile partitioning, and modulate the outgoing radiative flux, thereby shaping planetary cooling. At the same time, reducing conditions can favour the formation of thick H$_2$-rich atmospheres that, at sufficiently high pressures, also enhance greenhouse blanketing and prolong magma-ocean cooling times \citep{pierrehumbert_2011, lichtenberg_vertically_2021, nicholls_magma_2024}. When volatile species such as sulphur are included, the redox state further controls the emergence of H-C versus C-N-S-dominated atmospheric compositions \citep{gaillard_redox_2022}.

However, most existing models consider volatile inventories dominated by H$_2$O and CO$_2$ \citep{lebrun_thermal_2013, salvador_relative_2017, hier_2017, wordsworth_redox_2018, nikolaou_what_2019, barth_2021} H$_2$O \citep{hamano_2015, schaefer_predictions_2016}, or H$_2$O, CO$_2$, and O$_2$ \citep{krissansen-totton_oxygen_2021}. In parallel, coupled frameworks linking radiative–convective atmospheres with vertically resolved mantle evolution have also been developed \citep{lichtenberg_vertically_2021}. Their model included seven atmospheric species (H$_2$, H$_2$O, CO$_2$, CH$_4$, CO, O$_2$, and N$_2$) and simulated the coupled evolution of the magma ocean solidification, and the resulting atmospheric spectral signatures. From these simulations, \citet{lichtenberg_vertically_2021} showed that atmospheric composition strongly controls cooling timescales, with H$_2$-dominated atmospheres producing the longest-lived magma oceans.

More recently, fully coupled models have emerged that include the regulation of the atmospheric inventory through feedbacks between the interior and atmosphere, and compute volatile outgassing and atmospheric composition during magma-ocean evolution. These feedbacks arise from the coupled thermal evolution, whereby interior melting and degassing influence atmospheric opacity and cooling, which in turn affect the rate of interior solidification \citep{krissansen-totton_oxygen_2021, joshua_2024, nicholls_magma_2024}, these studies were compiled and reviewed by \citet{Lichtenberg2025TrGeo,Lichtenberg2025arXiv}.

Collectively, these studies demonstrate that the oxidation state of the magma ocean strongly controls planetary thermal evolution and atmospheric composition. As long as the magma ocean remains in equilibrium with the atmosphere and the mantle volatile reservoir greatly exceeds that of the atmosphere, the interior composition effectively buffers the atmospheric redox state. Recent work \citep{walbecq_2025} suggests that outgassing may be limited by dynamical constraints on volatile exsolution and transport, so interior–atmosphere equilibrium is not guaranteed during magma ocean solidification. Delayed degassing can therefore accelerate cooling and alter both atmospheric composition and mantle redox evolution. Regardless of outgassing efficiency, the mantle’s baseline redox state is inherited from high-pressure differentiation, set by equilibrium between metallic iron and iron oxide \citep{wade_2005, schlichting_chemical_2022, young_earth_2023, schaefer_ferric_2024, cottrell_2025}. Core-formation models further indicate that Earth’s mantle ended accretion in a relatively reduced state, around IW–2.0 i.e. two log units above the iron–wüstite buffer \citep{badro_core_2015, rubie2015formation}.

Nevertheless, following \citet{hirschmann_magma_2012}, recent experiments and molecular dynamics simulations suggested that high pressures may stabilize ferric (Fe$^{3+}$) over ferrous (Fe$^{2+}$) iron in silicate melts \citep{zhang_effect_2017, armstrong_deep_2019, deng_magma_2020, kuwahara_hadean_2023, schaefer_ferric_2024}. As a result, silicate melts equilibrated at depth may contain non-negligible fractions of ferric iron even under globally reducing conditions. During core formation, the segregation of metallic iron implies that a well-mixed mantle inherits the elevated Fe$^{3+}$/Fe$^{2+}$ ratios established at high pressure, resulting in higher effective oxidation states at lower pressures \citep{hirschmann_magma_2022, schaefer_ferric_2024}.

Chemical partitioning during magma ocean evolution further modifies mantle oxidation. Fractional crystallisation enriches FeO in the residual melt \citep{elkins-tanton_magma_2003, elkins-tanton_linked_2008, boukare_timing_2018,maurice_small-scale_2024}, potentially forming a dense, iron-rich layer that can trigger mantle overturn \citep{elkins-tanton_magma_2003, boukare_2015,boukare_timing_2018}. Because ferric iron (Fe$^{3+}$) is moderately incompatible relative to ferrous iron (Fe$^{2+}$) \citep{canil_distribution_1996, sorbadere_behaviour_2018}, crystallisation drives their fractionation \citep{mccanta_expanding_2009}. This redistribution establishes evolving redox gradients, linking interior differentiation to the oxidation state of surface volatiles \citep{hirschmann_magma_2012, sossi_2020}.

Beyond its role in volatile and redox exchange, \fotwo also directly modifies the melting behaviour of the mantle. In this work, we distinguished between two physically distinct oxidation states: a deep-mantle oxygen fugacity, \fotwomelt, which governs melting behaviour at high pressure, and a surface oxygen fugacity, \fotwosurf, which controls near-surface outgassing and volatile speciation. This distinction reflects an effective pressure-dependence of $f$O$_2$, consistent with experimental and theoretical results indicating that oxidation state may vary between deep and shallow mantle conditions. Recent experiments by \citet{lin_melting_2024} suggested that increasing \fotwomelt depresses the solidus relative to reduced baseline measurements \citep{tronnes_2000, tronnes_2002,ishii_phase_2018}. This effect is attributed to increased Fe$^{3+}$ concentrations in the melt, which act as incompatible species that lower the activities of major cation-bearing components and thermodynamically stabilize the liquid phase relative to crystalline minerals. As a result, oxidizing conditions can significantly depress the solidus by hundreds of Kelvin across a broad pressure range highlighting the combined influence of melt composition and pressure on magma ocean stability \citep{shahar_2021, lin_melting_2024}. This redox-dependent shift in the melting curves influences where a planetary interior temperature profile intersects the solidus, thereby affecting the onset of crystallisation, the depth and longevity of the magma ocean, and the efficiency of volatile exchange between the interior and the growing atmosphere. In principle, increasing \fotwo should shift melting curves to lower temperatures, allowing melting to persist to greater depths and potentially prolonging magma-ocean lifetimes, whereas more reducing conditions would be expected to favour earlier crystallisation.

In addition to the effects of oxygen fugacity, dissolved volatiles also modify mantle melting relations. Experimental studies indicate that H$_2$O and CO$_2$ can significantly depress silicate melting temperatures by altering melt structure and phase relations. In particular, water promotes hydroxylation of the silicate network \citep{katz_2003}, whereby H$_2$O is incorporated into the melt as OH$^-$, weakening the mineral framework and stabilizing the liquid phase. This volatile-induced modification of melt properties lowers the solidus and can influence the thermal evolution and longevity of magma oceans \citep{dasgupta_2007, myhill_2017, xie_crystallization_2024}. However, explicitly resolving how evolving volatile contents feedback on mantle melting introduces additional complexity that lies beyond the scope of the present study.

In this work, we extended the \texttt{PROTEUS} simulation framework \citep{lichtenberg_vertically_2021, nicholls_magma_2024, nicholls_self-limited_2025, nicholls_volatile-rich_2025, nicholls_thesis_2026} to quantify the influence of redox-dependent melting relations and volatile speciation on the coupled thermal and compositional evolution of magma ocean planets. We explicitly distinguished between the deep-mantle oxidation state, \fotwomelt, which governs melting behaviour and interior cooling, and the surface oxidation state, \fotwosurf, which regulates outgassing and atmospheric composition. As a reference case, we adopted the super-Earth exoplanet GJ~1132~b, discovered by \citet{berta_2015} using the MEarth-South telescope array. It orbits a nearby M dwarf star at 12.04~pc, with a mass of $1.62 \pm 0.55~M_\oplus$, a radius of $1.16 \pm 0.11~R_\oplus$, and a bulk density of $6.0 \pm 2.5$~g\,cm$^{-3}$, consistent with a rocky composition \citep{southworth_2017}. Receiving approximately 19 times Earth's insolation, it represents a benchmark target for testing magma ocean and atmospheric evolution models under strong stellar irradiation. With $T_{\mathrm{eq}} \approx 584$\,K and $A_{\rm B} = 0.19^{+0.12}_{-0.15}$\citep{xue_2024}, its atmospheric characterisation history includes previously claimed detections of H$_2$O, CH$_4$ and H$_2$-dominated envelopes \citep{southworth_2017, swain_2021, may_2023}. More recent and higher-precision JWST observations have progressively converged toward a bare-rock interpretation \citep{xue_2024, bennett_additional_2025}. We selected GJ~1132~b as our reference case because it combines well-characterised planetary parameters, favourable observational accessibility, and a stringent set of atmospheric constraints that can directly test the evolutionary pathways predicted by our models. Using this system, we investigated how variations in deep-mantle and surface redox reservoirs control mantle cooling history, solidification timescales, and volatile release. By systematically exploring different combinations of \fotwomelt and \fotwosurf together with the initial hydrogen inventory ($H_{\mathrm{oceans}}$) inherited from accretion, we aim to quantify the geophysical conditions under which redox processes may leave detectable imprints on the secondary atmospheres of super-Earth exoplanets.

\section{Methods}
\label{methods}
\begin{figure*}[t!]
    \centering
    \includegraphics[width=\textwidth,
                     keepaspectratio]{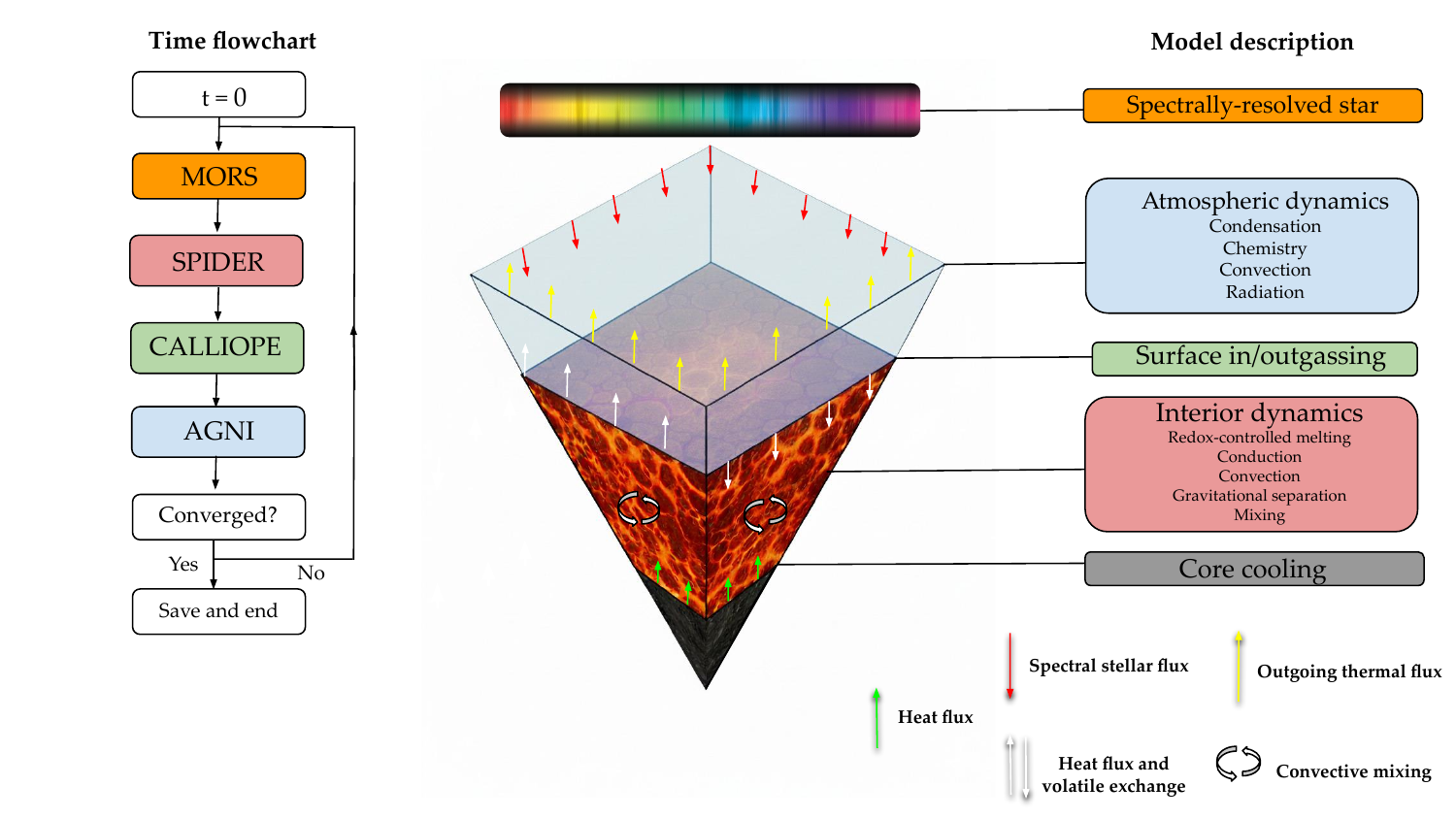}
    \caption{Schematic overview of the \texttt{PROTEUS} framework and its time-integration workflow. The interior, surface volatile exchange, and radiative--convective atmosphere are solved as independent modules that iteratively exchange boundary conditions at each timestep. In the time-flowchart, each timestep proceeds by iterating the coupled modules until the surface fluxes, volatile inventories, and atmospheric structure converge to a self-consistent solution. If convergence is not achieved, the boundary conditions are iteratively updated until convergence of the coupled solution is reached. The colour-coding in the time-flowchart (left) corresponds to the module descriptions in the model diagram (right), where each coloured module in the flowchart solves the physical processes listed in its corresponding coloured box on the right.}
    \label{fig:proteus_flowchart}
\end{figure*}

\subsection{\texttt{PROTEUS} overview}
To simulate the geophysical and atmospheric evolution of magma ocean exoplanets, we upgraded and employed the \texttt{PROTEUS} framework\footnote{\url{https://proteus-framework.org}} in version 25.07.31 \citep{lichtenberg_vertically_2021,nicholls_magma_2024,nicholls_self-limited_2025,nicholls_volatile-rich_2025}, a modular, self-consistent 1-D evolution code designed to track the coupled thermal and compositional history of rocky planets. \texttt{PROTEUS} is structured as a set of stand-alone but interdependent physical modules that simulates: interior, atmospheric energy transport, escape to space, stellar evolution, and outgassing that communicate through shared boundary conditions \citep{nicholls_thesis_2026}. At each interior timestep, energy, mass, and compositional fluxes are iteratively exchanged between modules until convergence of the coupled solution is achieved. The interior evolves forward in time, while the atmosphere is solved in quasi-steady state and responds instantaneously to evolving boundary conditions. This enables the model to track planets from an initially molten state through crystallisation and volatile exchange between mantle and atmosphere, resolving equilibrium states at each timestep and the hysteresis that emerges from coupled atmospheric–interior feedbacks \citep{Lichtenberg2025Sci,Lichtenberg2025arXiv, nicholls_self-limited_2025}. 

In this work, we advanced simulations until one of several termination criteria is met: (i) whole-mantle melt fraction decreases below a critical threshold $\phi$ < 0.05, indicating solidification; (ii) the system reaches global energy balance, where the melt fraction becomes constant in time and the net radiative flux between the interior-atmosphere is negligible ($\lesssim$ 1$~\mathrm{W\,m^{-2}}$); or (iii) a maximum integration time is reached (set to 200 Myr). This upper limit was chosen because the key phases of magma ocean evolution, including rapid crystallisation and convergence toward long-lived equilibrium states, occur within the first $10$-$100$~Myr \citep{nicholls_magma_2024}. Our focus is therefore on the early thermochemical evolution of the planet, and we do not attempt to model the subsequent long-term solid-state or tectonic evolution beyond this stage.

Figure~\ref{fig:proteus_flowchart} summarizes this workflow by showing how \texttt{PROTEUS} cycles through its modular components until convergence is achieved at each time step. In this study, we primarily modified the interior module (\texttt{SPIDER}) to incorporate redox-dependent melting relations, while the atmospheric, stellar evolution and volatile exchange modules follow previous implementations and are therefore only briefly described below.

\subsection{Stellar evolution model}

Within the \texttt{PROTEUS} framework, the temporal evolution of stellar irradiation is modelled using the \texttt{MORS} stellar evolution code, originally developed by \citet{johnstone_2021}. \texttt{MORS} models stars as coupled rotating shells surrounding a homogeneous core, allowing it to self-consistently calculate angular momentum transport and rotational spin-down over time. Since stellar activity and high-energy emission are closely linked to rotation, \texttt{MORS} then calculates the evolution of X-ray, extreme ultraviolet, and bolometric luminosities as a function of stellar mass, age, and initial rotation rate \citep{johnstone_2021}. These outputs are combined with empirical scaling relations to build the star’s evolving emission spectrum \citep{nicholls_magma_2024}. The resulting time-dependent stellar fluxes are continuously updated and coupled to the interior and atmospheric modules of \texttt{PROTEUS}, enabling a self-consistent treatment of how changing stellar irradiation influences planetary climate, chemistry, and interior–atmosphere feedbacks throughout the planet’s evolution \citep{tsai_2021, pierrehumbert_2010}. In this case, we used the star with a 50th percentile of rotation speed for its mass and age \citep{gallet}.

\subsection{Atmospheric model}

The energy transport in the atmosphere is computed with the open-source code \texttt{AGNI} \citep{nicholls_convective_2025, nicholls_volatile-rich_2025, nicholls_agni_2025}, a radiative–convective model that calculates the net atmospheric energy flux as a function of volatile composition, instellation flux, and the temperature at the atmosphere–mantle boundary \citep{nicholls_volatile-rich_2025}.
We employed \texttt{AGNI} in a configuration such that the atmospheric temperature–pressure profiles are convective (adiabatic). Based on this structure, \texttt{AGNI} calculates spectrally resolved radiative fluxes using the \texttt{SOCRATES} code under the two-stream approximation \citep{edwards_studies_1996,amundsen_uk_2016, sergeev_simulations_2023}, with gas opacities from the \texttt{DACE} \citep{grimm_2015, grimm_helios-k_2021} database and collision-induced absorption parametrised using \texttt{HITRAN} \citep{gordon_2022}.  
We assumed the atmosphere is isochemical at the outgassed composition (see below, Sect.~\ref{sec:outgassing}), with eight main volatile species: H$_2$O, H$_2$, CO$_2$, CO, SO$_2$, S$_2$, N$_2$, and CH$_4$. We assumed that the total number of moles of each volatile element is conserved throughout the simulation and do not model atmospheric escape. Consistent with our maximum integration time of 200 Myr, we focused on the early coupled interior–atmosphere evolution and do not attempt to model long-term atmospheric loss or the subsequent evolution toward present-day conditions. The impact of this assumption is discussed in Sect.~\ref{sec:discussion}.

\subsection{Outgassing model}
\label{sec:outgassing}

The exchange of volatiles between the molten mantle and the atmosphere is handled by the \texttt{CALLIOPE} outgassing module, based on \citep{bower_retention_2022}. This module applies mass conservation to solve for the C-H-O-N-S volatile partial surface pressures at thermochemical-solubility equilibrium at the magma ocean surface \citep{ chase_nistjanaf_1996,  bower_linking_2019,shorttle_distinguishing_2024, nicholls_magma_2024}. The corresponding abundances are computed based on their solubility and equilibrium between the melt phase and the atmosphere. During the evolution and without considering atmospheric escape, we conserved C-H-N-S and partitioned them between melt and atmosphere using solubility laws that depend on surface temperature, melt fraction, and imposed \fotwo \citep{Bower2025ApJ}.
Oxygen is not tracked as a finite reservoir; instead \fotwo is prescribed (infinite-reservoir approximation), so the atmosphere may effectively draw on or return O to the mantle to satisfy stoichiometry \citep{bower_retention_2022}.

\subsection{Interior model}
\label{sec:interior}
The planetary interior is modelled using the \texttt{SPIDER} code \citep{bower_numerical_2018, bower_linking_2019}, which simulates the thermal evolution of a magma ocean as it transitions from a fully molten state to partial or complete solidification. The model solves the vertically resolved energy balance using an entropy-based formulation of the conservation equation. The full numerical interior framework and implementation are described in detail in \citet{bower_numerical_2018}, and have been adopted in previous \texttt{PROTEUS} studies. Here, we briefly summarise the key aspects of the interior model, as our work focuses on how interior thermodynamics and melting behaviour influence atmospheric evolution.

In this context, the thermal evolution of the mantle is governed by the integral form of the energy conservation equation expressed in terms of entropy:

\begin{equation}
\int_V \rho\, T \,\frac{\partial S}{\partial t}\,\mathrm{d}V
= - \int_A \mathbf{F}\!\cdot\!\mathbf{n}\,\mathrm{d}A
+ \int_V \rho\, H\, \mathrm{d}V
\label{eq:energy_balance}
\end{equation}

\noindent with $S$ specific entropy, $\rho$ density, $T$ temperature, $F$ heat flux vector, $n$ unit normal vector to the surface, $H$ internal heat generation per unit mass, $t$ time, and $V$ volume.  The model takes into account conduction, mixing, gravitational separation and convection as energy transport mechanism, which contribute to the total flux $F$. Convective heat flux is computed through mixing length theory, enabling this mechanism to be represented as a diffusive 1D  process \citep{Wagner2019Apr}. In this work, we adopted a constant mixing length parametrization, following the framework explored in \citet{bower_numerical_2018}. In this way, the eddy diffusivity ($\kappa_h$) scales with a characteristic velocity depending on the local Reynolds number \citep{abe_1995,stothers_1997}, and affects the convective heat flux:
\begin{equation}
F_{\text{conv}} = - \rho T \kappa_h \frac{\partial S}{\partial r}.
\end{equation}
Energy transport contribution due to gravitational separation only occurs in the mixed phase region, by permeable flow of melt from the solid background, and depends on the flow mechanism (crystal settling or flotation), 
\begin{equation}
F_{\text{grav}} = 
a_g^2 \, g \, \rho \, \left( \rho_{\text{liq}} - \rho_{\text{sol}} \right)
\, \zeta_{\text{grav}}(\phi) \, T \, \Delta S_{\text{fus}} \, / \, \eta_m.
\end{equation}
where $a_g$ is the grain size, $\eta_m$ is the melt viscosity, $\rho_{\mathrm{liq}}$ and $\rho_{\mathrm{sol}}$ are the melt and solid densities at the liquidus and solidus, respectively, $\Delta S_{\mathrm{fus}} = S_{\mathrm{liq}} - S_{\mathrm{sol}}$ is the entropy of fusion defined as the difference between liquid and solid entropies and $\zeta_{\mathrm{grav}}(\phi)$ which parametrizes the dependence of the flow law on melt fraction $\phi$, and accounts for the transition from crystal-rich suspensions at high melt fractions to porous flow at low melt fractions \citep{abe_1988, bower_numerical_2018}.

Phase mixing quantifies the energy transport carried by the latent heat which is released/absorbed as crystals form/melt while they are being transported through the solidifying mantle,
\begin{equation}
F_{\text{mix}} = 
- \rho T \, \Delta S_{\text{fus}} \, \kappa_h \, \frac{\partial \phi}{\partial r},
\end{equation}
We considered a single-component system with two phases (solid and liquid), such that the melting region is represented by a finite interval between the solidus and liquidus. In \texttt{SPIDER}, these melting curves are implemented through entropy–pressure lookup tables that define the boundaries of this two-phase region. Because \texttt{SPIDER} solves the conservation of energy in terms of specific entropy, all thermodynamic quantities including temperature are obtained from tables \(T(S,P)\) and associated properties. At each pressure, the liquidus and solidus entropies \(\bigl(S_{\mathrm{l}}(P),\, S_{\mathrm{s}}(P)\bigr)\) determine the local melt fraction, which is computed as a linear function of entropy between these bounds (Eq.~\ref{eq:melt_fraction}). This representation allows \texttt{SPIDER} to flexibly incorporate melting relations from experimental petrology (e.g., peridotite and chondritic solidus/liquidus data) and to smoothly transition material properties across phase boundaries using numerical smoothing routines \citep{bower_numerical_2018}. Therefore, the lookup tables encode the full thermodynamic structure of the melting region and directly determine where the mantle is solid, partially molten, or fully liquid such that

\begin{equation}
\phi =
\begin{cases}
1, & \text{if } S > S_l \\
\dfrac{S - S_s}{S_l - S_s}, & \text{if } S_l \geq S \geq S_s \\
0, & \text{if } S_s > S
\end{cases}
\label{eq:melt_fraction}
\end{equation}

Conduction is described by Fourier's law, with thermal diffusivity $\kappa_t$, specific heat capacity $c$ and the adiabatic temperature gradient $(\partial T/\partial r)_S$, and is expressed as
\begin{equation}
F_{\text{cond}} = 
- \rho T \kappa_t \left( \frac{\partial S}{\partial r} \right)
- \rho c \kappa_t \left( \frac{\partial T}{\partial r} \right)_{S}.
\end{equation}
The total heat flux is expressed as the sum of the contributions of each heat flux associated to heat transport processes during the transition of a molten mantle:
\begin{equation}
F_{\mathrm{total}}= F_{\mathrm{conv}} + F_{\mathrm{mix}} + F_{\mathrm{cond}} + F_{\mathrm{grav}}.
\end{equation}

At each grid cell $F_{\mathrm{total}}$ can be positive or negative, defining the cooling or heating trajectory of individual layers within the planet's interior. Therefore, each of the fluxes will have a fundamental contribution to the total heat budget of the planet. 

We do not directly model the geophysical evolution of the metallic core, and we assumed the core-mantle boundary (CMB) to be thermally coupled to the mantle based on energy balance such that
\begin{equation}
\frac{\mathrm{d}T_{\text{CMB}}}{\mathrm{d}t} =
- \frac{4 \pi r_{\text{CMB}}^{2} F_{\text{CMB}}}
{m_{\text{core}} c_{\text{core}} \hat{T}_{\text{core}}}
\end{equation}
where $m_{\mathrm{core}}$, $c_{\mathrm{core}}$, $\hat{T}_{\text{core}}$, and $F_{\mathrm{CMB}}$ are the mass, heat capacity, thermal structure correction factor of the core, and heat flux at the CMB, respectively \citep{bower_numerical_2018}. The core heat capacity $c_{\mathrm{core}}$ is prescribed as a constant value representative of liquid iron (880~$\mathrm{J\,kg^{-1}\,K^{-1}}$), assuming an isentropic, well-mixed core.

Material properties are computed using separate equations of state for the melt and solid phases. The MgSiO$_{3}$ melt is described using the \texttt{RTpress} equation of state \citep{wolf_equation_2018}, which spans a wide pressure--entropy ($P$--$S$) range and is calibrated to experimental data using a limited set of physical parameters. For the solid phase, we adopted the MgSiO$_3$ equation of state from \citet{mosenfelder_mgsio3_2009}.  For the mixed-phase region, the properties are computed according to the local melt fraction Eq.~\eqref{eq:melt_fraction}, and assuming volume additivity when computing the two-phase adiabat \citep{solomatov_2007}. In this case, the melting curves employed are joint data to cover the entire mantle pressure range. At high pressures, the solidus and liquidus are derived from experiments on synthetic chondritic compositions \citep{andrault_solidus_2011}, while at lower pressures the melting relations follow the peridotite-based parametrization compiled by \citet{hamano_emergence_2013}. These datasets are smoothly joined to provide continuous melting curves across the mantle.

\subsection{Effect of $f\mathrm{O}_{2}$ on mantle solidus}
\label{sec:fo2_dep}

Building on the thermodynamic framework provided by the adopted equations of state, we now describe how variations in oxygen fugacity influence mantle melting relations. The melting temperature of the mantle depends sensitively on its composition and thermodynamic state, which are strongly regulated by redox conditions. Laboratory measurements by \citet{lin_melting_2024} showed a strong dependence of the solidus on \fotwomelt under oxidizing conditions (IW+2.0) at pressures of 16-26 GPa. Comparing these results with previous experiments under more reducing conditions \citep{tronnes_2000, tronnes_2002, ishii_phase_2018}, they find a shift of several log units in \fotwomelt between the two regimes. This behaviour can be expressed as:
\begin{equation}
T^{f\mathrm{O}_{2}}_{\text{solidus}} \, (\pm 383\mathrm{K})
= T^{\mathrm{IW+2.0}}_{\text{solidus}}
+ \frac{340}{3.2} \, \Delta \log f_{\rm{O_{2}}},
\label{eq:lin+24}
\end{equation}

With $\Delta$log\fotwo being the difference in log units between IW+2.0 (reference solidus) and the target \fotwo,and the $\pm383$\,K term reflects the 
experimental uncertainty on the solidus temperature derived from the 
\citet{lin_melting_2024} measurements. This would imply that very reducing mantles like the early Earth ($f$O$_2 \approx$ IW–4.0) should have a significantly higher melting temperature, whereas more oxidizing conditions lead to a lower melting temperature. This suggests that melting processes, redox evolution, outgassing, atmospheric composition and greenhouse effects form an emergent, coupled system that cannot be treated piecewise in isolation. To investigate these coupled interactions, we applied the \citet{lin_melting_2024} redox-dependent shift (Equation~\ref{eq:lin+24}) to the baseline melting curves used in our thermophysical model (Section \ref{sec:interior}).

\begin{figure}[h]
    \centering
    \includegraphics[width=\linewidth]{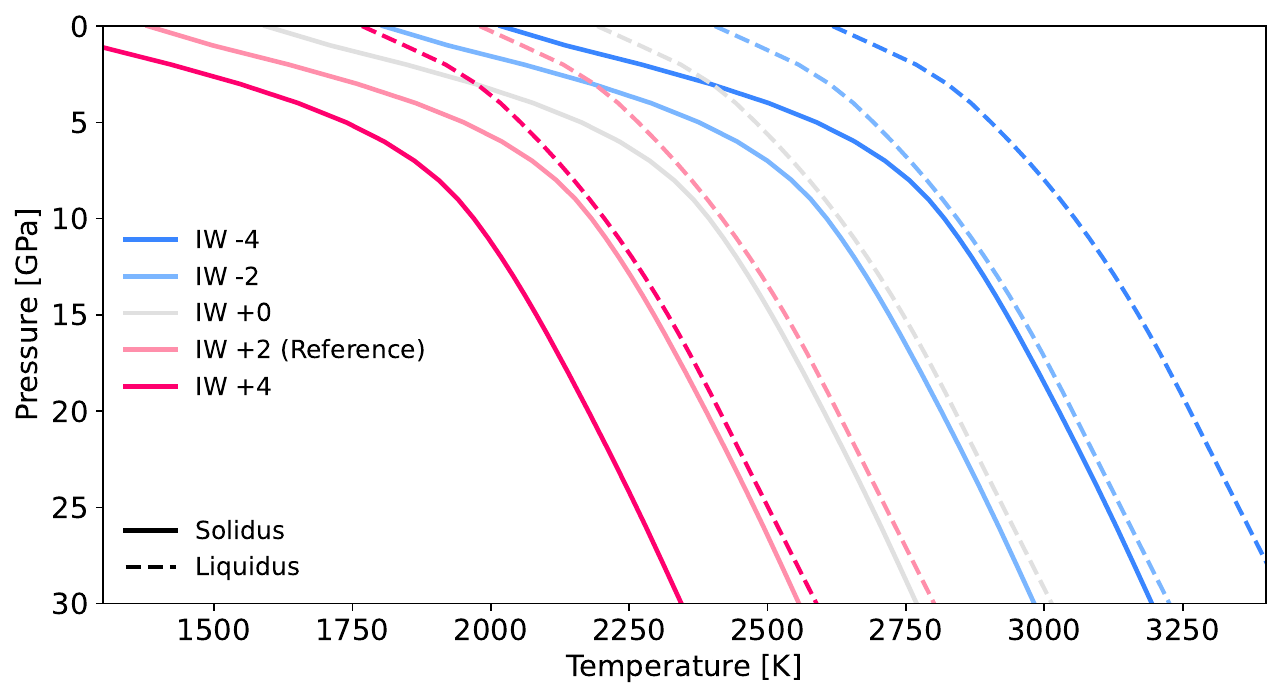}
    \caption{Solidus (solid lines) and liquidus (dashed lines) for mantle redox states relative to the iron--w\"ustite buffer (IW-4.0, IW-2.0, IW+0, IW+2.0, IW+4.0). Colours indicate 
    increasing oxidation from dark blue (IW-4.0) to pink (IW+4.0). The reference melting curves correspond to IW+2.0, representing an oxidized, broadly Earth-like mantle state consistent with experimental constraints \citep{andrault_solidus_2011,hamano_emergence_2013}. Other curves are obtained by applying the redox-dependent parametrization of \citet{lin_melting_2024}. Curves are shown up to 30~GPa for visualisation, although the full melting curves extends to 400~GPa; the redox-dependent temperature offset remains constant across this pressure range.}
    \label{fig:liquidus_all}
\end{figure}

In this study, the applied offsets produce the expected monotonic trend: reducing conditions (IW–4.0) shift the melting curves to higher temperature, whereas oxidizing conditions (IW+4.0) shift them to lower temperature. Because robust parametrizations describing the redox dependence of the liquidus are currently unavailable due to the limited experimental constraints, we applied the same temperature offset derived for the solidus to the liquidus for each redox state. This preserves the relative separation between solidus and liquidus while shifting the melting interval consistently with redox conditions. The resulting set of redox-dependent melting curves is shown in Fig.~\ref{fig:liquidus_all}.

In our framework, these redox-dependent melting curves provide a convenient parametrization of melting behaviour as a function of redox state and are used directly within \texttt{SPIDER} to compute the melt fraction. We noted, however, that oxygen fugacity is fundamentally an outcome of thermodynamic equilibrium rather than an independently imposed parameter. During magma ocean evolution, melting induces phase separation and element partitioning, which can modify the Fe$^{3+}$/Fe$^{2+}$ ratio and hence the effective $f$O$_2$, leading to a corresponding evolution of the melting relations themselves. In this work, we therefore adopted a simplified approach in which redox-dependent offsets to the melting curves are prescribed, and should be interpreted as a first-order approximation to this coupled thermochemical behaviour.

Because the oxidation state of the mantle is likely inhomogeneous with pressure, we incorporated two distinct effects of $f\mathrm{O}_{2}$ that reflect two different physical roles of redox state during magma–atmosphere evolution. In both cases, we explored the same range of redox conditions (IW–4.0, IW–2.0, IW+0, IW+2.0, IW+4.0), allowing us to connect interior melting behaviour with surface outgassing. The individual effects of these two $f\mathrm{O}_{2}$ terms are detailed as follows:
\begin{enumerate}
    \item Deep-mantle oxidation state \fotwomelt controlling the melting curves. \fotwomelt enters our model through the redox‐dependent shift of the melting curves, as described in Section \ref{sec:fo2_dep}; we applied the chosen value of $f$O$_{2}$ to adjust the melting curve used by the interior model. This \fotwomelt value expresses the thermophysical and compositional state of the interior.
    \item Surface oxidation state \fotwosurf governing outgassing and volatile speciation. We parametrized \fotwosurf to express the oxidation state at the surface. This value is used in the \texttt{CALLIOPE} module, where it controls volatile speciation and the composition of the outgassed atmosphere. Due to near-surface processes, in particular the pressure dependence of $f$O$_2$, \fotwosurf may differ by orders of magnitude from \fotwomelt. This behaviour is consistent with Earth, where near-surface reservoirs maintain oxidation states substantially higher than those of the bulk mantle \citep{cottrell_2025}.
\end{enumerate}

Varying \fotwosurf and \fotwomelt independently enables us to investigate how mantle redox conditions shape atmospheric composition during the magma-ocean stage. By allowing interior and outgassing $f\mathrm{O}_{2}$ values to vary separately, the model provides a first-order approach for exploring how potential redox gradients between the deep mantle and the magma ocean surface influence thermal and chemical evolution.

\subsection{Parameter space exploration}

To test the effects of the varying redox states on the geophysical and melting evolution of young rocky planets, we ran a grid of simulations for the specific case of the super-Earth exoplanet GJ 1132 b, where we fixed the melting curve at different redox states to study the possible implications on the subsequent formation of the atmosphere and the evolution of mantle melt fraction. We also explored the orbital distance of exoplanets as exerting significant control over the stellar irradiation, which strongly affects the total heat flux to space. We probed a temperate and highly irradiated case. The orbital separation $a$ was set by defining two reference instellations at 4.5 Gyr for an Earth-like planet, such that our simulated planets receive (i) the stellar irradiation as the Earth today at an age of 4.5 Gyr, and (ii) the same as the real exoplanet GJ 1132 b.

Primordial volatile inventories and bulk compositions of close-in rocky exoplanets remain poorly constrained, motivating an exploration of a wide range of plausible initial conditions. We therefore varied both the initial C/H ratio and the total hydrogen abundance across a broad parameter space in units of Earth oceans. Our nominal values are guided by estimates for the early Earth \citep{wang_elemental_2018}, whereas oxygen fugacity is varied across a range motivated by expected mineralogical diversity, which alone may generate differences of at least four log units in $f\mathrm{O}_2$ \citep{guimond_mineralogical_2023}, extending from highly reduced Mercury-like conditions \citep{cartier_role_2019} to more oxidized scenarios like the Earth's mantle today \citep{stagno_2020}. In all simulations, we adopted a fixed core radius fraction of 0.55 $R_{\rm p}$
 and an initial surface temperature of 3300~K, chosen to ensure that the initial adiabat lies above the liquidus throughout the mantle. This approach allows us to assess the sensitivity of magma ocean evolution and volatile outgassing to poorly constrained initial compositions. The complete simulation grid comprises 600 models and is summarized in Table~\ref{tab:params}.

\begin{table}[t!]
\centering
\caption{Overview of the parameter space explored in this study.}
\label{tab:params}
\begin{tabular}{ll}
\hline\hline
Parameter & Values \\
\hline
\fotwomelt, relative to IW+2 & $-4.0,\,-2.0,\,0.0,\,+2.0,\,+4.0$ \\
\fotwosurf & $-4.0,\,-2.0,\,0.0,\,+2.0,\,+4.0$ \\
Total planet mass [$M_\oplus$] & $1.0,\,3.0$ \\
Semimajor axis [au] & $0.0153,\,0.0700$ \\
Bulk inventory [$H_{\mathrm{oceans}}$] & $3,\,10,\,100$\\
Bulk C/H mass ratio & $0.1,\,2.0$ \\
\hline
\end{tabular}
\end{table}

\section{Results}
\subsection{Effect of redox-dependent melting curves on thermal evolution}
\label{sec:results_overview}

\begin{figure*}[t!]
    \centering
    \includegraphics[width=\linewidth]{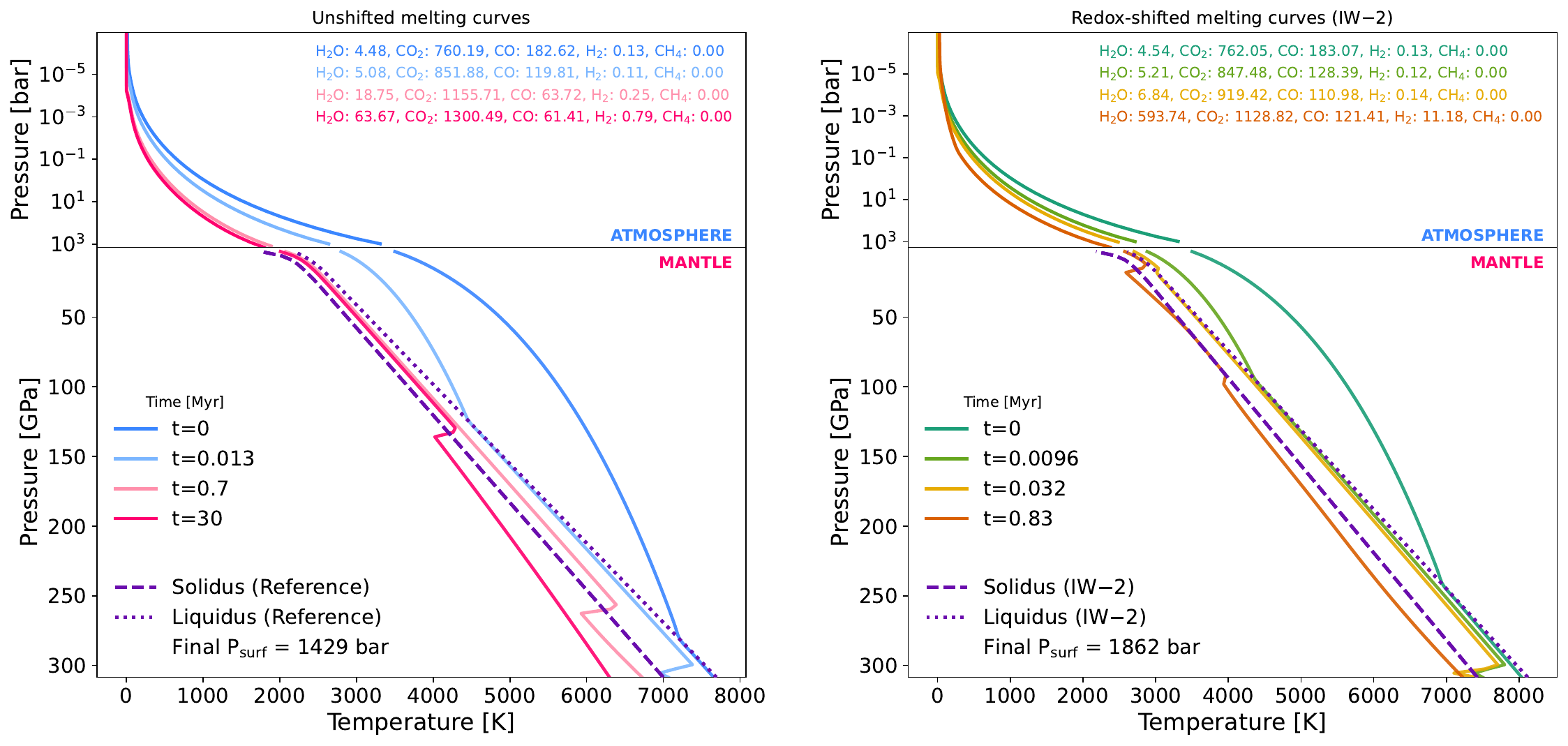}
    \caption{$T$–$P$ profiles for two simulations of a 3~$M_\oplus$ planet showing the coupled interior–atmosphere evolution from an initially global magma ocean to radiative equilibrium (partial pressures in bar).
Left: Baseline melting curves from \citet{andrault_solidus_2011,hamano_emergence_2013}. The mantle remains partially molten for an extended period and solidifies through fractional crystallisation, reaching steady state after $\sim$30~Myr with a surface pressure of $\approx$1429~bar.
Right: Redox-dependent melting curves shifted to higher temperatures (IW-2.0). Earlier deep-mantle crystallisation confines melt to a shallow magma ocean sustained by greenhouse warming from outgassed water. Steady state is reached after 0.83~Myr with a surface pressure of $\approx$1862~bar.}
    \label{im:profiles_unaffected}
\end{figure*}

To begin with, we explored how the redox-dependent melting behaviour introduced in Section~\ref{sec:fo2_dep} affects the thermal evolution of the planetary interior during magma ocean evolution. Fig ~\ref{im:profiles_unaffected} compares two representative simulations with an initial hydrogen inventory of $3~H_{\mathrm{oceans}}$, computed under an instellation equal to Earth's present-day incident flux, which differ only in their choice of melting curves: the unshifted experimentally based curves from \citet{andrault_solidus_2011} and \citet{ hamano_emergence_2013}, and the same curves shifted 
according to the redox-dependent parametrization of \citet{lin_melting_2024} 
assuming the mantle is buffered at IW+2.0 relative to the iron--w\"ustite buffer. As a result of the differing melting curves, the thermal evolution is not a simple rescaling of temperature: shifts in the melting region produce a markedly non-linear response in the coupled interior–atmosphere system.

In the baseline case, a cooler set of melting curves allows a substantial region of the mantle to remain partially molten over an extended period. The melt is relatively well distributed throughout the interior, and the planet cools through a long-lived phase in which deep melt continues to supply heat to the surface. In contrast, the shifted IW-2.0 melting curves favours earlier and more extensive crystallisation at depth. As the rheological front moves upward, melt becomes increasingly confined to a thin shell near the surface, so that the late-time configuration is dominated by a shallow surface magma ocean rather than a deeply distributed molten layer.

In both simulations, we followed the evolution until they reach a steady state, but the internal melt distribution and atmospheric structure differ substantially. In both cases, the thermal profile rapidly approaches the rheological transition, which sets the bulk melt fraction and thereby regulates volatile partitioning between the interior and atmosphere. In our simulations, the rheological front marks the depth at which the effective viscosity increases abruptly from $\sim10^{2}$ to $\sim10^{22}$~Pa~s, separating low-viscosity convecting melt from a high-viscosity, partially crystallized layer. This transition arises from the viscosity parametrization as a function of melt fraction, with a sharp increase occurring around a critical melt fraction of $\sim$40\%, where the solid phase starts to control the dynamic timescale \citep{bower_retention_2022}. In the IW-2.0 case, water initially dissolved in the deep magma is progressively exsolved as crystallisation proceeds, increasing the atmospheric opacity and enhancing greenhouse warming. This feedback allows the thin, near-surface magma layer to persist even after most of the deeper mantle has solidified. The comparison demonstrates that an \fotwomelt-dependent shift of the melting curves does not simply modify the overall thermal state, but instead shifts the reference point relative to the rheological transition. This alters the position of the rheological front and, consequently, the timescales and volatile evolution pathways, determining whether the system evolves toward a deep, partially molten interior or a shallow surface magma ocean.

\subsection{Impact of parameters on the interior properties}
\label{impact_physical}

\begin{figure*}[htb] 
    \centering
    \includegraphics[width=\textwidth]{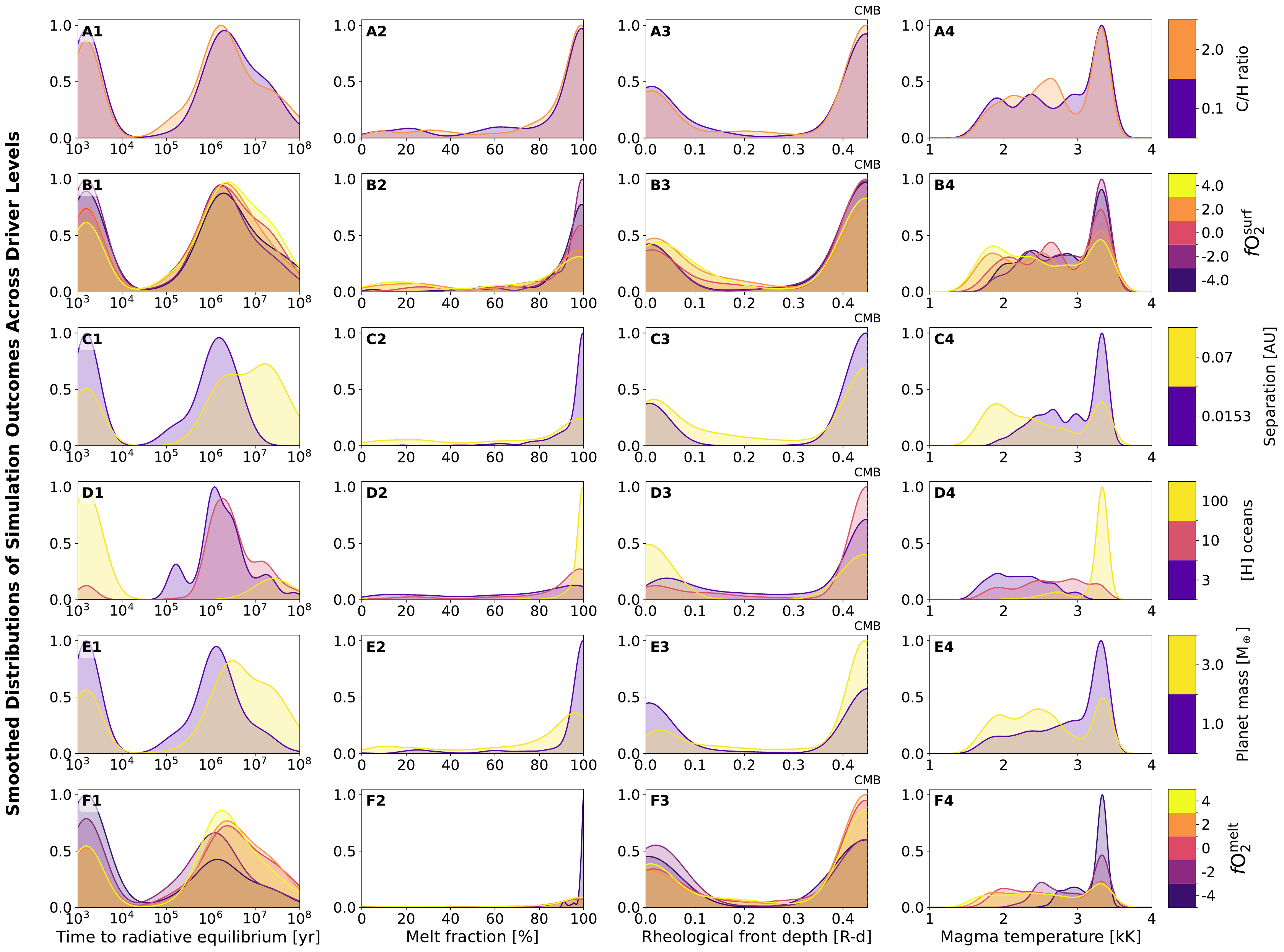}
\caption{Smoothed distributions of simulation outcomes across the parameter grid.
Rows show variations in the driving parameters (C/H ratio, surface oxidation state \fotwosurf relative to the iron–wüstite buffer, orbital separation, volatile inventory in H-ocean mass, planetary mass, and deep-mantle oxidation state \fotwomelt), while columns show interior outcomes (time to radiative equilibrium, bulk melt fraction, rheological-front depth, and surface temperature).
Coloured curves represent kernel density estimates of simulations sharing a given parameter value. This smoothing is used only for visualisation to highlight trends and clustering across the discrete simulation outcomes.}
    \label{grid_physical}
\end{figure*}

As a next step, we characterised the diversity of final states and magma ocean lifetimes by examining the global structure of the parameter grid. We analysed a suite of 505 out of 600 simulations that successfully ran to completion, where the 600 models result from the combination of all parameter values listed in Table~\ref{tab:params}. The remaining 95 simulations did not converge, predominantly at the most reducing melting curve (IW$-$4.0), where the elevated solidus temperatures led to numerical instabilities at high pressures. All converged simulations reached radiative equilibrium before fully solidifying, indicating the presence of long-lived magma oceans of varying sizes. Regarding the choice of 200~Myr integration time, as noted in Section~\ref{methods}, the key phases of magma ocean evolution occur within the first tens to hundreds of of megayears \citep{nicholls_magma_2024}. The converged simulations all reach radiative equilibrium well within this timescale, suggesting that extending the cutoff further would not significantly affect our results. Fig ~\ref{grid_physical} shows the distribution of interior properties as a function of the main physical and compositional drivers, with rows representing the independent parameters (C/H ratio, surface redox state log(\fotwosurf/IW), orbital separation, hydrogen inventory, planet mass, and deep-mantle oxidation state \fotwomelt) and columns corresponding to key outcome variables (solidification time, melt fraction, rheological-front depth, and near-surface surface temperature). The density distributions demonstrate that not all regions of parameter space contribute equally: some parameter combinations cluster around distinct evolutionary pathways, while others produce more extreme or less frequent outcomes.

Two overarching features emerged from these distributions. First, the rheological front depth (Fig.\ref{grid_physical}, third column; panels A3–F3) exhibits a pronounced bi modality, with peaks near the core–mantle boundary and close to the surface. 
This bi modality reflects the existence of two dominant evolutionary pathways (first column; panels A1–F1): planets that reach radiative equilibrium at very early stages while remaining fully molten, characterized by near-unity melt fractions and rheological fronts anchored at the core-mantle boundary (located at $\sim 0.55$ in normalized planetary radius, corresponding to the adopted core radius fraction), and planets that undergo partial crystallisation near the surface, leading to reduced melt fractions and shallow viscosity contrasts. To complement the density distributions (Fig~\ref{grid_physical}), we additionally showed cumulative distributions (Fig~\ref{fig:grid_impactogram} in Appendix) to quantify the fraction of the simulation ensemble that occupies each regime and to verify that the inferred shifts are systematic rather than driven by low-probability tails.

Among the individual driver parameters, the deep-mantle oxidation state \fotwomelt emerges as an important parameter for the melt reservoir through redox-dependent shifts (Fig~\ref{grid_physical} row F). Reduced melting curves systematically favour earlier interior crystallisation and higher surface temperatures, whereas oxidized melting curves maintain lower melt fractions and delay crystallisation. The surface oxidation state \fotwosurf exerts a primary control on surface temperature through its impact on volatile speciation affecting atmospheric composition and the consequent radiative feedbacks, producing modest shifts toward lower temperatures under more oxidized surface conditions (panel B4).

\begin{figure*}[htb] 
    \centering

    \includegraphics[width=0.99\textwidth]{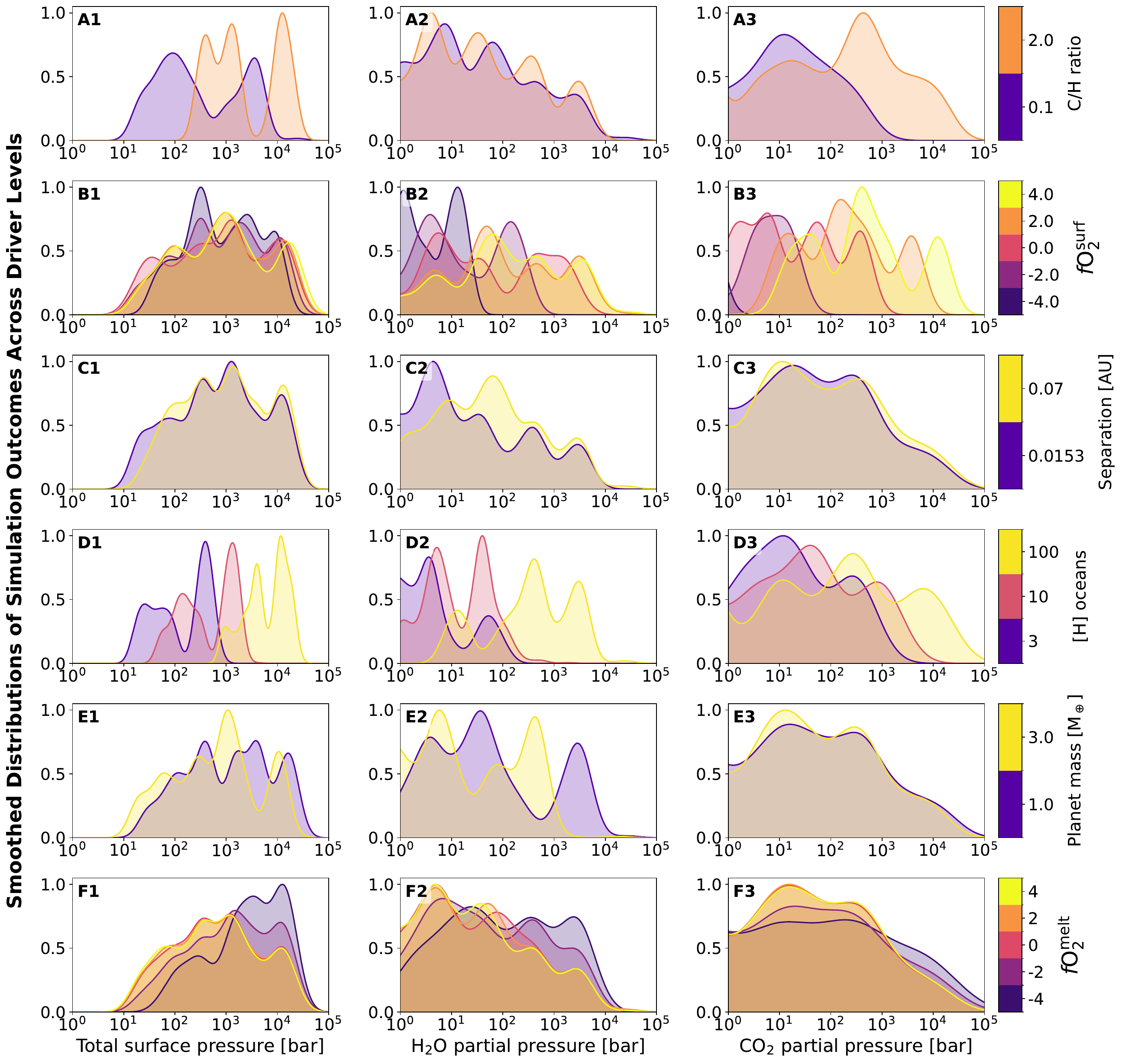}
    \caption{Smoothed density distributions of atmospheric outcomes across the parameter grid.
Rows show variations in the driving parameters (C/H ratio, surface oxidation state \fotwosurf relative to the iron–wüstite buffer, orbital separation, volatile inventory in H-ocean mass, planetary mass, and deep-mantle oxidation state \fotwomelt), while columns show interior outcomes (time to radiative equilibrium, bulk melt fraction, rheological-front depth, and surface temperature).
Coloured curves represent kernel density estimates of simulations sharing a given parameter value. This smoothing is used only for visualisation to highlight trends and clustering across the discrete simulation outcomes.}
    \label{grid_atmo}
\end{figure*}

Initial volatile inventories, particularly hydrogen abundance (row D), further regulate surface temperature and melt persistence by controlling greenhouse blanketing and the timing at which radiative equilibrium is established. Large $H_{\mathrm{oceans}}$ drive a substantial fraction of simulations toward near 100\% melt fractions (panel D2) and higher surface temperatures (panel D4), because volatile-rich planets can reach radiative equilibrium early while remaining fully molten. In this volatile-rich limit, the evolution of the rheological front is therefore primarily set by atmospheric feedbacks. As a consequence, \fotwomelt exerts a reduced influence on the evolution, and \fotwosurf instead controls both the total atmospheric mass and its speciation. Variations in the C/H ratio further modulate this behaviour (Fig \ref{grid_atmo} panel A1), as carbon-bearing species exhibit low solubility and preferentially partition into the atmosphere, contributing to the bi modality observed in rheological front regimes.

Structural and orbital parameters introduce additional but subdominant effects (row C). Close-in and low mass planets populate hotter regimes, while more distant and more massive planets preferentially occupy cooler states. These trends mainly redistribute the population along the surface temperature without altering the underlying bimodal structure.

Taken together, the combined density distributions and cumulative statistics demonstrate that magma-ocean evolution is governed by a hierarchy of controls. The volatile inventory acts as the primary regulator, determining whether \fotwomelt can modulate the melt reservoir size and consequently the atmospheric feedbacks that set the surface temperature. Orbital separation and planet mass play subordinate roles within the explored parameter space.

\subsection{Impact of parameters on atmospheric properties}

Similarly, Fig~\ref{grid_atmo} and \ref{cumulative_atmo} shows the distribution of atmospheric outcomes across the full parameter grid as a function of the main physical and compositional drivers in the \texttt{PROTEUS} simulations. Each column represents a key atmospheric quantity, including total surface pressure, and the H$_{2}$O and CO$_{2}$ partial pressure. The overlapping probability and cumulative distributions provide a sensitivity map of how different input parameters regulate volatile outgassing and atmospheric composition.

Initially, the bulk C/H ratio distributions primarily reflect the melt state established by volatile-rich
cases that reach radiative equilibrium early in their evolution. Within this framework, variations in C/H modulate how carbon is partitioned between the interior and the atmosphere. Because carbon is only weakly soluble in silicate melt, simulations that remain largely molten preferentially retain carbon-bearing species in the atmosphere, leading to enhanced surface pressures in carbon-rich systems. These effects are compounded by carbon’s high molecular weight, which further enhances surface pressure in C-dominated atmospheres, consistent with the results shown in \citet{bower_retention_2022}. 

The surface oxidation state \fotwosurf exerts a primary control on regulating atmospheric speciation. Variations in \fotwosurf produce important changes in H$_{2}$O and CO$_{2}$ partial pressures. Orbital separation shows only a weak influence on atmospheric outcomes (row C). While more distant planets tend to occupy slightly cooler regimes, the surface and partial pressures remain broadly similar across the explored range. Therefore, during the magma ocean stage, volatile in/outgassing are primarily controlled by internal thermochemical processes rather than external irradiation, for these two modestly-irradiated scenarios. The initial hydrogen inventory strongly regulates atmospheric mass and thermal state through greenhouse feedbacks. Large volatile budgets (10--100 $H_{\mathrm{oceans}}$) drive a substantial fraction of simulations toward high surface pressures, near-unity melt fractions, and elevated surface temperatures. 

Planetary mass (row E) introduces only minor variations in atmospheric outcomes across the explored parameter space. In our model, mass enters the interior structure through hydrostatic equilibrium, where it determines the radial pressure profile and thus the mantle grid, and through the gravitational acceleration ($g$), which controls the pressure–depth relation and influences volatile solubility and degassing. In the atmosphere, $g$ further affects the pressure structure and radiative properties. Despite these dependencies, the distributions of total atmospheric mass, surface pressure, and volatile partial pressures remain broadly similar for the 1~$M_\oplus$ and 3~$M_\oplus$ cases, indicating that gravitational and structural effects play a limited role in regulating outgassing in these scenarios.

Finally, the deep-mantle oxidation state \fotwomelt (row F) affects atmospheric properties through its control on mantle melt reservoir size as mentioned in Section \ref{impact_physical}, therefore showing higher total surface pressures for reducing \fotwomelt, due to early crystallisation and outgassing of volatiles.

\subsection{Impact of \fotwomelt on atmospheric composition}

To examine in more detail how \fotwomelt and volatile inventory shape the chemical structure of the outgassed atmospheres, we next analysed the final surface pressure and volatile speciation as a function of \fotwomelt and hydrogen inventory quantified in Earth oceans. While the impactograms in Fig~\ref{grid_atmo} summarize global sensitivities across the parameter space, the following figures resolve individual volatile contributions and quantify how different species partition into the atmosphere under contrasting redox and inventory regimes. This allows us to directly link deep-mantle oxidation state and volatile budgets to observable atmospheric compositions.

\begin{figure*}[htb]
    \centering
    \includegraphics[width=\linewidth]{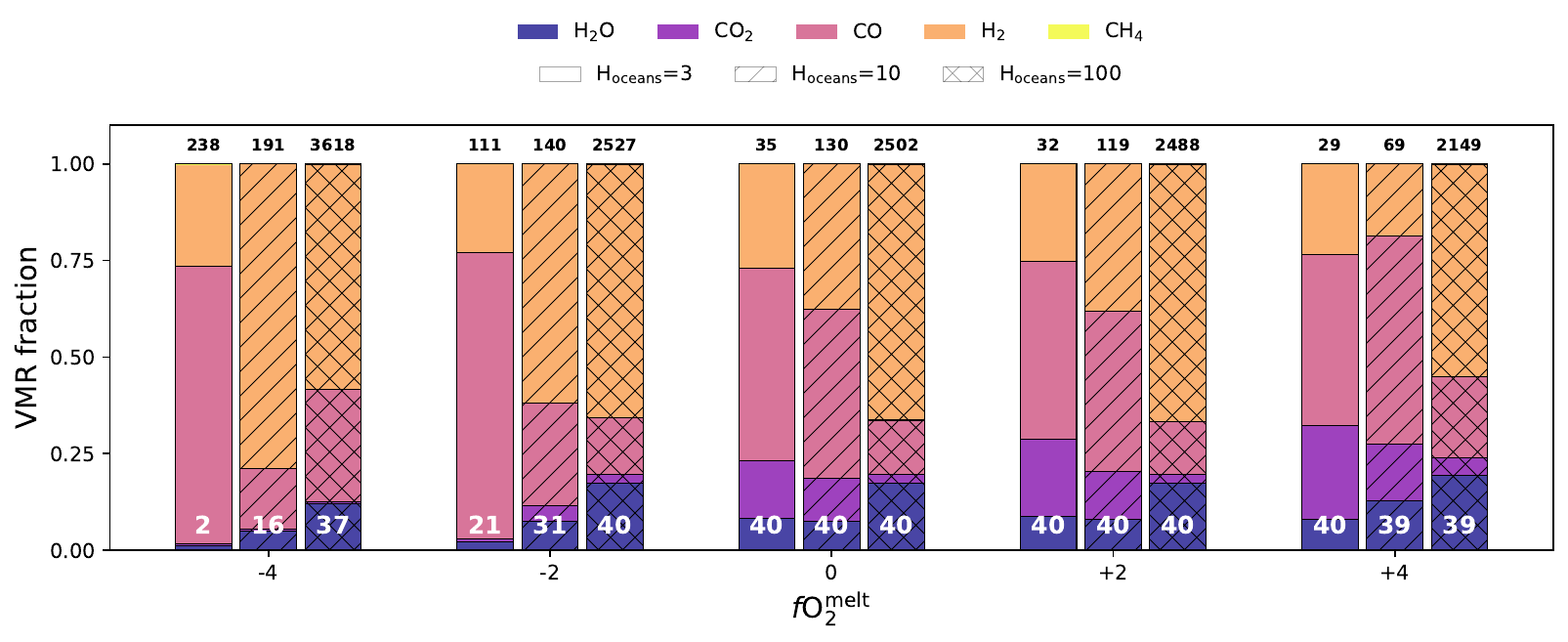}
    \caption{Volume mixing ratio of outgassed volatiles as a function of the
deep-mantle oxidation state \fotwomelt for simulations with initial water
inventories of 3, 10, and 100~$H_{\mathrm{oceans}}$. Numbers within the 
bars indicate the number of simulations contributing to each median value 
and numbers on top of each bar represent the median surface pressure 
$P_{\mathrm{surf}}$ in bar. Colours denote the relative contributions of 
major atmospheric species (H$_2$O, CO$_2$, CO, H$_2$, CH$_4$), while 
hatch patterns distinguish different volatile inventories. CH$_4$ is 
present in individual simulations but contributes negligibly to the median 
volume mixing ratio and is therefore not visible in the stacked bars. 
The low number of converged simulations for $H_{\mathrm{oceans}} = 3$ at 
\fotwomelt $= \mathrm{IW}-4$ ($n=2$) reflects numerical instabilities 
arising from the elevated solidus temperatures under strongly reducing 
conditions, which prevented convergence in most of these runs.}

    \label{im:histogram}
\end{figure*}

Figure~\ref{im:histogram} shows the median final surface pressure ($P_{\mathrm{surf}}$) and atmospheric composition for simulations with 3, 10, and 100~ $H_{\mathrm{oceans}}$. Bars are stacked by volatile species (H$_2$O, CO$_2$, CO, H$_2$, CH$_4$), illustrating the relative contribution of each gas to the total pressure. 

The results reveal a systematic decrease in total surface pressure with increasing deep-mantle oxidation state \fotwomelt, consistent across all volatile inventories. Under strongly reducing melting conditions (\fotwomelt = IW-4.0 to IW–2.0), the atmospheres are thick and dominated by H$_2$ and CO, with pressures reaching 238~bar for 3~ $H_{\mathrm{oceans}}$ and exceeding 3500~bar for 100~$H_{\mathrm{oceans}}$. As \fotwomelt becomes more oxidizing ($\Delta$IW > 0) along the melting curve, the total pressure drops sharply—by an order of magnitude in the low-volatile cases—while the relative abundance of oxidized species (CO$_2$ and H$_2$O) increases. This transition marks a \fotwosurf-controlled compositional switch from H$_2$-rich reducing atmospheres to CO/CO$_2$-rich oxidizing ones, reflecting equilibrium partitioning during magma ocean outgassing. In parallel, the initial hydrogen inventory also plays a major role: larger volatile budgets (10–100~$H_{\mathrm{oceans}}$) yield proportionally more massive atmospheres,consistent with expectations from equilibrium models
\citep{ sossi_2020,nicholls_magma_2024, shorttle_distinguishing_2024}.

\subsection{Impact of \fotwomelt on bulk melt fraction}

\begin{figure}[h]
    \centering
    \includegraphics[width=\linewidth]{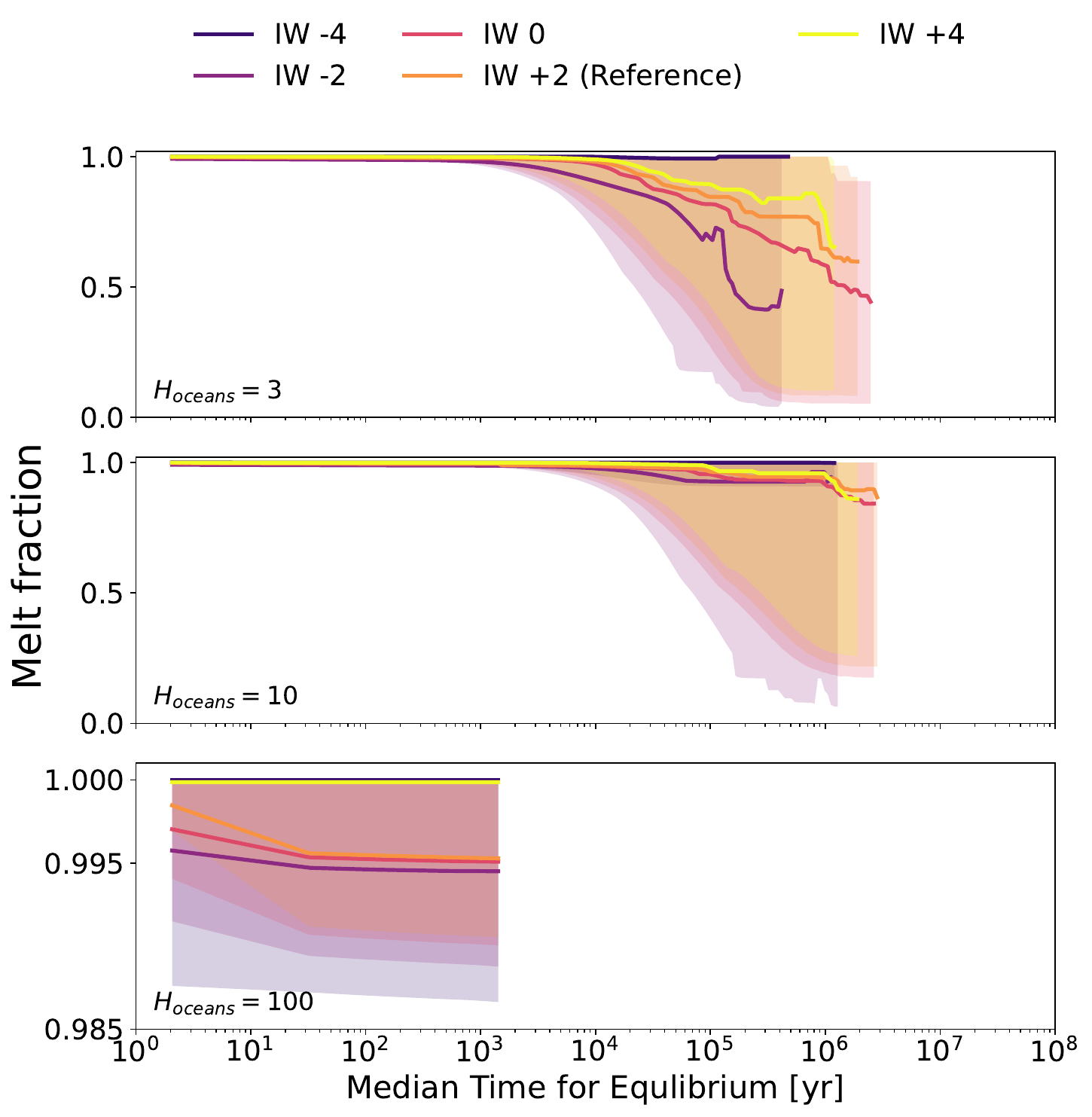}
    \caption{Evolution of the mantle melt fraction with time for three initial water inventories ($H_{\mathrm{oceans}}$ = 3, 10, 100; top to bottom) and different redox-dependent melting scenarios (\fotwomelt: IW–4.0 to IW+4.0). Curves show the median across simulations at each time, with shaded regions indicating the full range. Each curve is truncated at the median time for equilibrium.}
    \label{fig:melt_fraction_evolution}
\end{figure}

Additionally, Fig~\ref{fig:melt_fraction_evolution} further shows that the bulk melt fraction provides a unifying control on volatile partitioning by setting the effective size of the silicate reservoir available for dissolution as mentioned before. In volatile-rich cases ($H_{\mathrm{oceans}}$=100), the mantle remains nearly fully molten and reaches equilibrium rapidly, such that the reservoir size is both large and nearly constant in time. As a result, differences in \fotwomelt produce only minor variations in melt fraction and therefore have a negligible impact on volatile partitioning. In contrast, volatile-poor systems ($H_{\mathrm{oceans}}$=3) evolve over longer timescales and exhibit substantial changes in melt fraction, reflecting a progressively shrinking reservoir. Consequently, variations in \fotwomelt strongly affect the rate and extent of crystallization, and therefore directly control the mass of volatiles that can be retained in the interior versus released to the atmosphere. This behaviour is consistent with the trends identified in the impactogram ~\ref{fig:grid_impactogram}, and supports a simplified interpretation in which deep-mantle redox state governs the mass balance of volatile partitioning through its control on bulk melt fraction, while surface redox conditions regulate the chemical speciation of the atmosphere.

Overall, these results demonstrate that the deep-mantle oxidation state \fotwomelt and the initial volatile inventory jointly regulate the timing and the amount of volatiles released by controlling how the bulk melt fraction evolves over time. As the interior progressively solidifies, volatiles are released at different rates, while the surface temperature further influences atmospheric composition and, in turn, the cooling rate. By controlling the mantle melting curves and therefore the melt fraction, \fotwomelt governs how long and how extensively the mantle remains molten, which in turn determines the capacity of the interior to exchange volatiles with the atmosphere. In contrast, the surface oxidation state \fotwosurf primarily controls atmospheric speciation by regulating near-surface chemical equilibria, determining whether outgassed volatiles are retained as H$_2$-rich or CO$_2$-dominated atmospheres. While higher water inventories mainly set the absolute atmospheric mass by sustaining large and persistent magma oceans, they do not systematically follow the redox-dependent melting trends imposed by \fotwomelt. Together with the distribution analysis in Fig~\ref{im:histogram}, these results show that the combined effects of deep-mantle melting behaviour and surface redox chemistry control the distribution of atmospheric outcomes—from thick, H$_2$-rich atmospheres to thinner, oxidized envelopes—across different redox states, primarily in volatile-poor systems. This coupling implies that the final atmospheric composition, and more indirectly the surface pressure, may retain signatures of the planet’s interior oxidation state, with observable molecular abundances providing the most robust diagnostic.

\subsection{Impact of \fotwomelt on interior-atmosphere partitioning}

\begin{figure*}[htb] 
    \centering
    \includegraphics[width=0.99\textwidth]{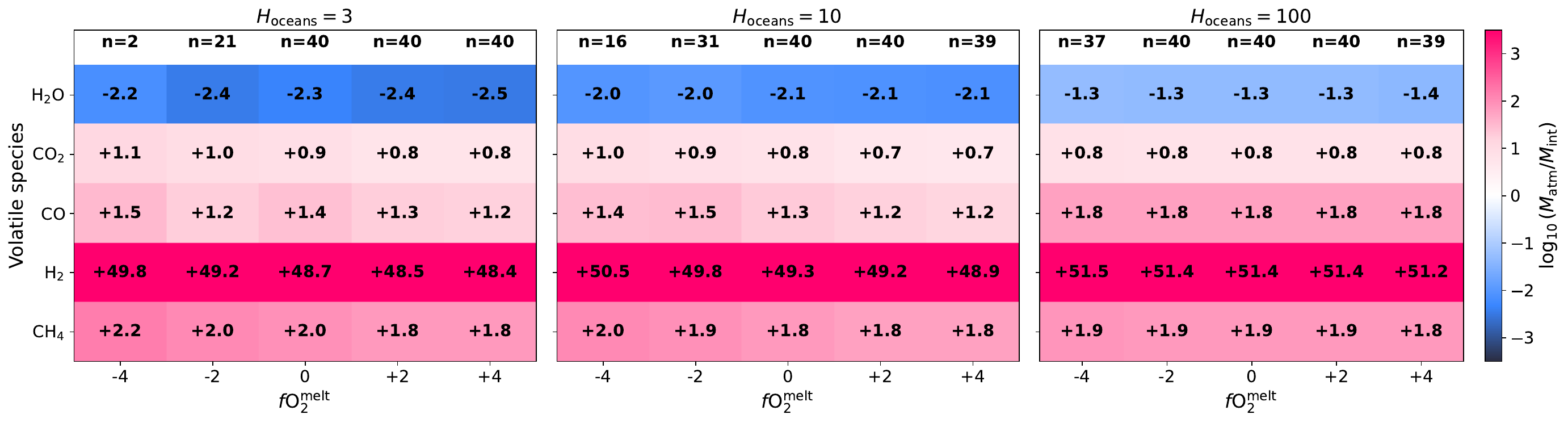}
\caption{Atmosphere-interior partitioning of major volatile species as a function of deep-mantle oxidation state \fotwomelt and initial volatile inventory. Each panel corresponds to a different initial H inventory (3, 10, and 100~$H_{\mathrm{oceans}}$). Colours indicate the median logarithmic ratio of atmospheric to interior volatile mass, $\log M_{\mathrm{atm}}/M_{\mathrm{int}}$, for each species and oxidation state offset along the melting curve (\fotwomelt, expressed relative to the IW+2.0). Positive values denote atmosphere-dominated partitioning, whereas negative values indicate retention within the mantle. Numbers above each column indicate the number of converged simulations contributing to each bin. The figure highlights three distinct regimes: persistent interior buffering of H$_2$O, systematic atmospheric dominance of H$_2$, and strong redox-dependent partitioning of carbon-bearing species.}

    \label{im:heatmap}
\end{figure*}

Finally, Fig~\ref{im:heatmap} summarizes the atmosphere-interior partitioning of major volatile species as a function of deep-mantle oxidation state \fotwomelt and initial volatile inventory. Each panel corresponds to simulations with 3, 10, and 100~$H_{\mathrm{oceans}}$, and colours indicate the median logarithmic ratio of atmospheric to interior volatile mass, $\log M_{\mathrm{atm}}/M_{\mathrm{int}}$, for each species and \fotwomelt. Positive values denote atmosphere-dominated partitioning, whereas negative values indicate preferential retention within the mantle. This representation provides a direct, quantitative measure of degassing efficiency and reveals how mantle redox conditions and volatile inventory jointly regulate volatile partitioning during magma ocean outgassing.

Across all simulations, H$_2$O remains strongly retained within the mantle, with median values of $\log M_{\rm atm}/M_{\rm int} \simeq -2$ to $-1$, indicating that only a few percent of the total H$_{2}$O inventory is transferred to the atmosphere. This behaviour reflects the high solubility of water in silicate melt and its limited degassing efficiency under high melt fractions \citep{bower_retention_2022, Bower2025ApJ}. In contrast, carbon-bearing species exhibit a pronounced behaviour with \fotwomelt. Both CO and CO$_2$ preferentially partition into the atmosphere, with typical ratios of $\log M_{\rm atm}/M_{\rm int} \simeq$ +1.5 to +0.8 under reducing conditions. CO$_2$ displays its highest atmosphere-interior ratios in reduced mantles and progressively lower values toward more oxidizing regimes, reflecting the redox-dependent melting behaviour imposed by \fotwomelt, with enhanced interior retention at high oxidation states due to high melt fractions.

Hydrogen-bearing species show the strongest partitioning contrast. Molecular hydrogen is highly atmosphere-dominated across all redox states, with $\log M_{\rm atm}/M_{\rm int} \approx 48$–52. These extreme values arise specifically for H$_2$, which is effectively insoluble in silicate melt and therefore absent from the interior \citep{sossi_solubility_2023}; for visualisation facilities, interior H$_2$ masses are represented using a minimum value of $10^{-30}$kg. Methane also preferentially partitions into the atmosphere, but exhibits a \fotwomelt-dependent behaviour similar to that of carbon-bearing species, with higher $\log M_{\rm atm}/M_{\rm int}$ ratios under reducing conditions and progressively lower values toward oxidizing regimes.

Comparison across volatile inventories indicates that volatile-poor systems often display more extreme $\log M_{\rm atm}/M_{\rm int}$ ratios, implying stronger preferential partitioning into either the atmosphere or the interior. In contrast, volatile-rich cases generally exhibit smaller ratios for the same species and redox conditions, reflecting a more balanced distribution of volatiles between the mantle and the atmosphere.  Overall, these results show that volatile partitioning is primarily controlled by the bulk melt fraction, which governs the efficiency of degassing. The melt fraction is, in turn, modulated by the deep-mantle oxidation state \fotwomelt through its influence on the melting curves. As a result, the impact of \fotwomelt on volatile partitioning is indirect and most pronounced in volatile-poor systems, whereas the initial volatile inventory mainly sets the absolute atmospheric mass. The chemical speciation of the outgassed volatiles is determined by near-surface conditions during degassing \citep{gaillard_2014}, rather than by deep-mantle redox alone.

\section{Discussion}
\label{sec:discussion}

A central question of this study is whether the \fotwomelt-dependent melting temperature differences can meaningfully influence planetary evolution. Our results show that the answer depends critically on the volatile inventory. Redox state undoubtedly affects the melting curve: reducing conditions shift it to higher temperatures, whereas oxidizing conditions lower it. This offset determines when the temperature profile first intersects the solidus and, consequently, the conditions under which solidification initiates. However, whether this difference propagates into large-scale evolutionary consequences depends on whether the thermal profile intersects the melting curves before the system approaches thermal equilibrium.

In volatile-rich cases ($H_{\mathrm{oceans}}$ = 10–100), the mantle thermal profile remains far above the solidus for most of the evolution. These simulations never reach the melting region at all, because the surface cools into radiative equilibrium long before the interior thermal profile approaches the melting point. As a result, even a several-hundred-Kelvin redox shift produces little effect on melt fraction, cooling timescales, or the depth of the convecting layer. In this regime, \fotwomelt does not strongly influence the planet’s thermal pathway. Instead, the \fotwosurf mainly controls the chemistry of the  atmosphere: reduced mantles yield H$_2$–CO-rich envelopes, whereas oxidized mantles favour CO$_2$–H$_2$O atmospheres. Differences in final atmospheric mass remain detectable across redox states, but they are not as large as in the volatile-poor case.

By contrast, in volatile-poor cases ($H_{\mathrm{oceans}}$ = 3), the interior 
thermal profile approaches and in some simulations intersects the melting region 
before the system reaches radiative equilibrium. Here, \fotwomelt plays a decisive 
role. Hotter melting curves associated with reduced conditions (IW-2.0) shift the melting region to higher temperatures, leading to earlier progression towards 
solidification at depth and reorganizing the distribution of melt within the 
planet. Instead of sustaining a deep magma ocean as in the reference case, 
reduced melting curves promote more extensive interior solidification, reducing 
the depth of the molten region and confining melt to a shallow near-surface layer. 
At early times, however, the migration of the rheological front is controlled by 
the atmospheric cooling rate, as the interior efficiently resupplies heat to the 
surface. In this configuration, water released during interior solidification 
increases atmospheric opacity and enhances greenhouse warming, thereby slowing 
cooling and allowing the shallow surface magma layer to persist. In contrast, 
cooler melting curves associated with oxidized conditions (IW+2.0 to IW+4.0) 
allow extensive melting to persist in the deep interior for longer, maintaining a deep and 
vertically extended magma ocean.

In this volatile-poor regime, therefore, \fotwomelt influences not only the timing of crystallisation but also the depth of the magma ocean, the distribution of melt available to host volatiles, and the efficiency of outgassing. These effects amplify differences in atmospheric pressure and composition across redox states, demonstrating that \fotwomelt influences both the structure of the mantle and the eventual atmospheric properties for this specific scenario.

\subsection{Implications for atmospheric composition and observability}

Across the full grid, reduced mantles (IW-4.0 to IW-2.0) produce $\mathrm{H_2}$-dominated atmospheres, with CO as the main carbon carrier. Oxidized mantles (IW+2.0 to IW+4.0) generate $\mathrm{CO_2, CO}$ and H$_2$O-rich envelopes. This reflects well-known \fotwosurf -control on volatile speciation: hydrogen and carbon remain volatile under reducing conditions, but oxidize into less volatile species that dissolve more readily in the melt \citep{sossi_2020, nicholls_magma_2024, Bower2025ApJ, Lichtenberg2025Sci}.

These compositional trends directly affect the atmospheric mean molecular weight and total surface pressure. Reduced cases generate the most massive atmospheres, often exceeding one kilobar for the volatile-rich cases, while oxidized cases generate lower-pressure but higher mean molecular weight (MMW) envelopes. The radiative consequences are notable: high pressure $\mathrm{H_2}$+CO atmospheres efficiently trap heat within the planet, through collision-induced absorption \citep{borystow_h2_2002}, potentially prolonging magma ocean lifetimes \citep{nicholls_magma_2024, joshua_2024}. Oxidized atmospheres, though IR-active, are less massive and offer a weaker blanketing effect at late times. 

This redox–volatility coupling offers a possible explanation for the atmospheric diversity now being revealed in close-in rocky planets around M dwarfs. Observations of 55~Cnc~e \citep{hu_secondary_2024} and TOI-561~b \citep{teske_thick_2025}, for example, indicate volatile-rich envelopes dominated by $\mathrm{CO}$ and $\mathrm{CO_2}$, broadly consistent with the oxidized to moderately reduced regimes identified here. However, other ultra-short period  exoplanets do not exhibit signs of an atmosphere \citep{Lichtenberg2025TrGeo}. The high surface pressures and mixed C–O chemistry of the scenarios inferred for these systems could reflect planets that partially solidified from initially reducing magma oceans and later oxidized through volatile loss or mantle crystallisation. Water storage in oxidized mantles has been predicted to affect the long-term bulk volatile composition of observed super-Earths \citep{Dorn2021ApJL}. Conversely, more reduced cases in our grid predict thick $\mathrm{H_2}$+CO atmospheres. Therefore, our simulations illustrate that redox trends inherited from accretion can affect the distribution of melting in the deep interior, and, by extension, the storage capacity of the mantle over Gyr timescales.

\subsection{Benchmarking case}

GJ~1132~b provides a particularly valuable benchmark for testing coupled interior–atmosphere evolution models, given the availability of high-precision observational constraints. Early ground-based transmission spectroscopy observations \citep{diamond_2018} already indicated a largely featureless spectrum, disfavouring H$_2$-rich atmospheres and suggesting either a high mean molecular weight atmosphere or absence of an atmosphere.  Similarly, recent JWST/NIRSpec transmission spectroscopy observations shows that the planet spectrum is best described by a flat continuum \citep{bennett_additional_2025}, consistent with either a very thin steam atmosphere or an atmosphere-free surface. Nevertheless, when combined with MIRI/LRS thermal emission measurements, these data favour the presence of a very thin atmosphere explained by a nearly bare rocky planet \citep{xue_2024}. In this way, forward modelling of the dayside emission spectrum rules out Earth-like atmospheres ($P \sim 1$~bar) containing even modest amounts of H$_2$O, excludes atmospheres of any thickness (10$^{-4}$-10$^{2}$~bar) with $\gtrsim$1\% CO$_2$, and disfavours thick, Venus-like envelopes ($P \gtrsim 100$~bar) with trace amounts of CO$_2$ or H$_2$O, indicating that GJ~1132~b most likely lacks a significant volatile atmosphere \citep{xue_2024}.

Within the context of our simulations, the apparent lack of a substantial atmosphere on GJ~1132~b is most consistent with volatile-poor initial conditions and an oxidized deep-mantle state \fotwomelt that limits sustained melt fractions and long-term volatile release. Reduced and volatile-rich evolutionary pathways in our models tend to produce thick H$_2$- and CO-rich atmospheres during early magma-ocean phases, which, although potentially short-lived due to escape, would be expected to leave behind more substantial secondary atmospheres. This interpretation is consistent with the evolutionary models of \citet{schaefer_predictions_2016}, who investigated the coupled magma ocean–atmosphere evolution of GJ~1132~b. They find that retaining a substantial water-rich atmosphere requires extreme initial volatile inventories $\gtrsim$5~wt\% (percent by weight) of H$_2$O, while most evolutionary pathways lead to tenuous atmospheres. In their models, photodissociation of water and preferential hydrogen escape result in the build-up of abiotic O$_2$, with the magma ocean acting as only a limited oxygen sink. As a consequence, the most common outcome is a thin, O$_2$-dominated atmosphere, unless a long-lived magma ocean or unusually high initial water content is maintained. The observational evidence for GJ~1132~b therefore favours an evolutionary history in which efficient early atmospheric loss and limited late-stage outgassing suppress the persistence of a detectable present-day atmosphere. Future detections of even weak CO$_2$- or H$_2$O-dominated envelopes would provide valuable constraints on the planet’s initial volatile inventory and redox state, highlighting the potential of GJ~1132~b as a probe of coupled interior–atmosphere evolution \citep{nicholls_volatile-rich_2025}.
A key question emerging from our results is whether the interior oxidation state can be constrained from atmospheric observations alone. As shown in Fig~\ref{im:histogram}, volatile inventory exerts a stronger control on total atmospheric mass but not on composition than redox state, while \fotwosurf primarily controls atmospheric speciation. This introduces a degeneracy between volatile content and oxidation state that is difficult to break from bulk atmospheric properties alone. In this context, accurate atmospheric modelling plays an important role in linking observable upper-atmosphere compositions to near-surface redox chemistry. Mixing processes and photochemical reactions can modify the vertical distribution of molecular species, potentially obscuring or preserving the redox signature of outgassed volatiles in the observable atmosphere \citep{pierrehumbert_2010}. Turbulent mixing and photochemical reactions can modify the vertical distribution of molecular species, potentially obscuring or preserving the redox signature of outgassed volatiles in the observable atmosphere \citep{shorttle_distinguishing_2024, nicholls_self-limited_2025}. While thermochemical equilibrium signatures set in the deep atmosphere may be propagated to observable pressure levels if mixing timescales are short compared to reaction timescales \citep{tsai_2023, nicholls_self-limited_2025}. It should thus be tested under which conditions disequilibrium atmospheric processes could obscure the outgassed volatile composition and thus redox state. Together, these considerations suggest that these kind of modelling, would represent a promising avenue for constraining deep-mantle redox conditions from upcoming JWST and ELT observations.
In this study, we assumed volatile exchange only between the melt and atmosphere. In reality, incorporation of $\mathrm{H}$, $\mathrm{C}$, or $\mathrm{O}$ through melt trapping or crystal inclusions \citep{sim_volatile_2024,joshua_2024} could increase the total outgassing efficiency and accelerate mantle oxidation, potentially yielding thinner atmospheres and shorter magma ocean lifetimes \citep{lebrun_thermal_2013, salvador_relative_2017, nikolaou_what_2019, bower_linking_2019}. In these cases, melt redistribution due to the effects described here would amplify the ability of the mantle to store or release volatiles \citep{hier_2017, sim_volatile_2024}. The feedbacks between these mechanisms should be studied, but are outside the scope of this work.

Furthermore, atmospheric escape processes are not included. In particular, close-in rocky planets are expected to experience efficient hydrodynamic and energy-limited escape of light species driven by stellar X-ray and extreme ultraviolet irradiation (e.g.,\cite{watson_1981,zahnle_1986,schaefer_predictions_2016,owen_2019}). As a consequence, the large H$_2$ abundances predicted under reducing conditions in our simulations likely represent upper limits on the long-term atmospheric hydrogen content \citep{schaefer_predictions_2016, nicholls_magma_2024}. However, by varying the initial hydrogen inventory across our model grid, we effectively explored a range of outcomes that may arise from different escape histories, providing a first-order sensitivity analysis of how atmospheric mass and composition respond to hydrogen loss. In reality, substantial hydrogen loss may occur on relatively short timescales, potentially reducing the persistence and observability of H$_2$-rich atmospheres \citep{schaefer_predictions_2016}. In particular, predictions concerning the absolute abundance, lifetime, and detectability of H$_2$-dominated atmospheres would benefit most from a self-consistent treatment of escape processes.

In addition, we do not explicitly resolve a depth-dependent oxygen fugacity structure. In the present formulation, the redox dependence of melting is implemented as a temperature shift relative to a reference solidus, such that variations in \fotwomelt produce a uniform offset while preserving the underlying 
pressure dependence of the melting curves. As a result, the adopted parametrisation captures first-order redox effects as a global shift in melting behaviour. If \fotwomelt were instead resolved as a function of depth, this would lead to local modifications of the solidus temperature at each pressure level, effectively 
producing a depth-dependent perturbation to the melting relations and potentially shifting the depth at which the interior adiabat intersects the solidus. Near-surface reservoirs are generally expected to be more oxidized than the deep interior. While our parameter grid explores all combinations as a broad sensitivity study, scenarios in which \fotwosurf exceeds \fotwomelt are the more geophysically realistic ones. Given the complexity of self-consistently coupling a depth-dependent redox profile to the melting parametrisation, we adopted a qualitative approach and treat \fotwomelt and \fotwosurf as independent parameters, allowing us to bracket the range of plausible interior-atmosphere redox configurations.

Beyond the effects of oxygen fugacity on the melting curves, dissolved water in the melt would further modify our results through two main mechanisms. First, as referenced in Section~\ref{sec:intro} and Section \ref{sec:fo2_dep} of this study, water depresses the solidus through hydroxylation of the silicate network \citep{katz_2003}, whereby H$_2$O is incorporated into the melt as OH$^-$, weakening mineral bonds and stabilizing the liquid phase, with volatile-saturated silicates exhibiting solidus depressions of up to 600$-$800~K \citep{dasgupta_2007, myhill_2017, xie_crystallization_2024}. This would tend to maintain higher melt fractions for longer, extending magma ocean lifetimes and affecting outgassing rates, particularly in the volatile-poor regime where the thermal profile approaches the solidus. Second, and not accounted for in our model, water significantly lowers melt viscosity, which would enhance convective heat transport through the mantle. In this way, a less viscous melt would increase the efficiency of convection, therefore accelerating cooling.

Despite these assumptions, the physical trends identified in this work remain robust. Reduced planets may present favourable observational signatures.
For low to intermediate volatile inventories ($H_{\mathrm{oceans}} = 10$)  and moderately reducing mantles ($\mathrm{IW} -3$ to $-2$), the resulting $\mathrm{H_2}$- and CO-rich atmospheres produce strong molecular absorption features in the 3–5~$\mu$m range accessible to JWST and future ELT instruments \citep{katyal_effect_2020, nicholls_convective_2025}. More extremely reduced and volatile-rich cases ($\mathrm{IW} \lesssim -4$, $H_{\mathrm{oceans}} = 100$)  instead generate very thick $\mathrm{H_2}$ envelopes that may be susceptible to atmospheric escape. In contrast, oxidized mantles produce $\mathrm{H_2O}$- and $\mathrm{CO, CO_2}$-dominated atmospheres. These compositional differences suggest that future observations may help constrain deep-interior redox states and volatile inventories through atmospheric spectra. Therefore, the initial volatile inventory emerges as the primary control on atmospheric mass, \fotwosurf on the volatile speciation, while the influence of the deep-mantle oxidation state \fotwomelt plays a secondary role that depends strongly on the available volatile budget. As recent JWST observations highlight, the atmospheric properties of close-in rocky exoplanets are not yet fully explained. Integrating interior redox evolution, volatile partitioning, water-dependent melting feedbacks and atmospheric escape will therefore be essential for interpreting upcoming detections of secondary atmospheres and for linking observable spectra to the geochemical state of rocky exoplanet interiors like in the case of GJ~1132~b, TOI-561~b, and TOI-6255~b.

\section{Conclusions}

We upgraded the coupled numerical framework \texttt{PROTEUS} to investigate how deep-mantle redox state (\fotwomelt) and volatile inventory jointly shape magma ocean evolution and atmospheric composition in rocky exoplanets, using GJ~1132~b as a reference case. Our approach incorporates \fotwo-dependent shifts in mantle melting curves, allowing us to assess how variations in redox state propagate into the thermal structure, melt fraction, and atmospheric composition. We explored a wide range of redox states (IW-4.0 to IW+4.0) and volatile inventories to determine under which conditions redox-dependent melting curves significantly influence interior evolution and potentially observable atmospheric properties.

Our results reveal a hierarchy of controls on magma ocean evolution, where volatile inventory sets a primary control on whether the redox-dependent melting relations modulate or not bulk melt fraction, and consequently the size of the melt reservoir available for volatile dissolution and outgassing. Our main conclusions are as follows:

\begin{itemize}

\item Redox-dependent shifts in the melting curves become important for volatile-poor planets where the mantle thermal profile approaches the solidus. In this regime, hotter melting curves associated with reducing mantle conditions trigger earlier crystallisation, promote a shallow-surface magma ocean and enhance interior–atmosphere partitioning, whereas cooler melting curves associated with oxidizing conditions prolong the fully molten state and promote a more radially distributed melt reservoir. In contrast, for volatile-rich planets, the mantle thermal profile remains far above the melting region and reaches radiative equilibrium before crystallisation occurs, making the thermal evolution in this regime largely insensitive to \fotwomelt.

\item While \fotwomelt modulates bulk melt fraction in the volatile-poor regime, the surface oxidation state \fotwosurf regulates near-surface chemical equilibria and determines the speciation of outgassed volatiles. Reduced surface conditions favour H$_2$+CO-rich atmospheres, whereas oxidized conditions produce H$_2$O+CO$_2$ dominated envelopes.

\item Orbital separation plays a secondary role compared to volatile inventory and redox state.
Within the parameter space explored for GJ~1132~b, variations in orbital distance produce relatively minor changes in atmospheric outcomes compared with those driven by volatile content or $\mathrm{IW}$.

\end{itemize}

Future extensions of this work will incorporate time-dependent redox evolution, solid-phase volatile trapping, and atmospheric escape to enable fully self-consistent predictions of thermal evolution and spectroscopic observables for rocky exoplanets.

\begin{acknowledgements}
The authors thank the anonymous referee for their constructive comments, which helped improve the manuscript. MS thanks Laura Kreidberg, Mara Attia, Rob Spaargaren and Christopher Boettner for useful comments and suggestions. This work was partially supported by the Branco Weiss Foundation, the Netherlands eScience Center (\texttt{PROTEUS} project, NLESC.OEC.2023.017), the Alfred P. Sloan Foundation (AEThER project, G202114194), NASA’s Nexus for Exoplanet System Science research coordination network (Alien Earths project, 80NSSC21K0593), and the NWO NWA-ORC PRELIFE Consortium (PRELIFE project, NWA.1630.23.013). HN acknowledges support from STFC grant UKRI1184. We also thank the Center for Information Technology of the University of Groningen for their support and for providing access to the Habrok high performance computing cluster.
\end{acknowledgements}

\bibliographystyle{aa}
\bibliography{references_ads}

\onecolumn
\appendix
\section*{Appendix}

\setcounter{figure}{0}
\renewcommand{\thefigure}{A.\arabic{figure}}

\begin{figure}[htb]
    \centering
    \includegraphics[width=0.95\textwidth]{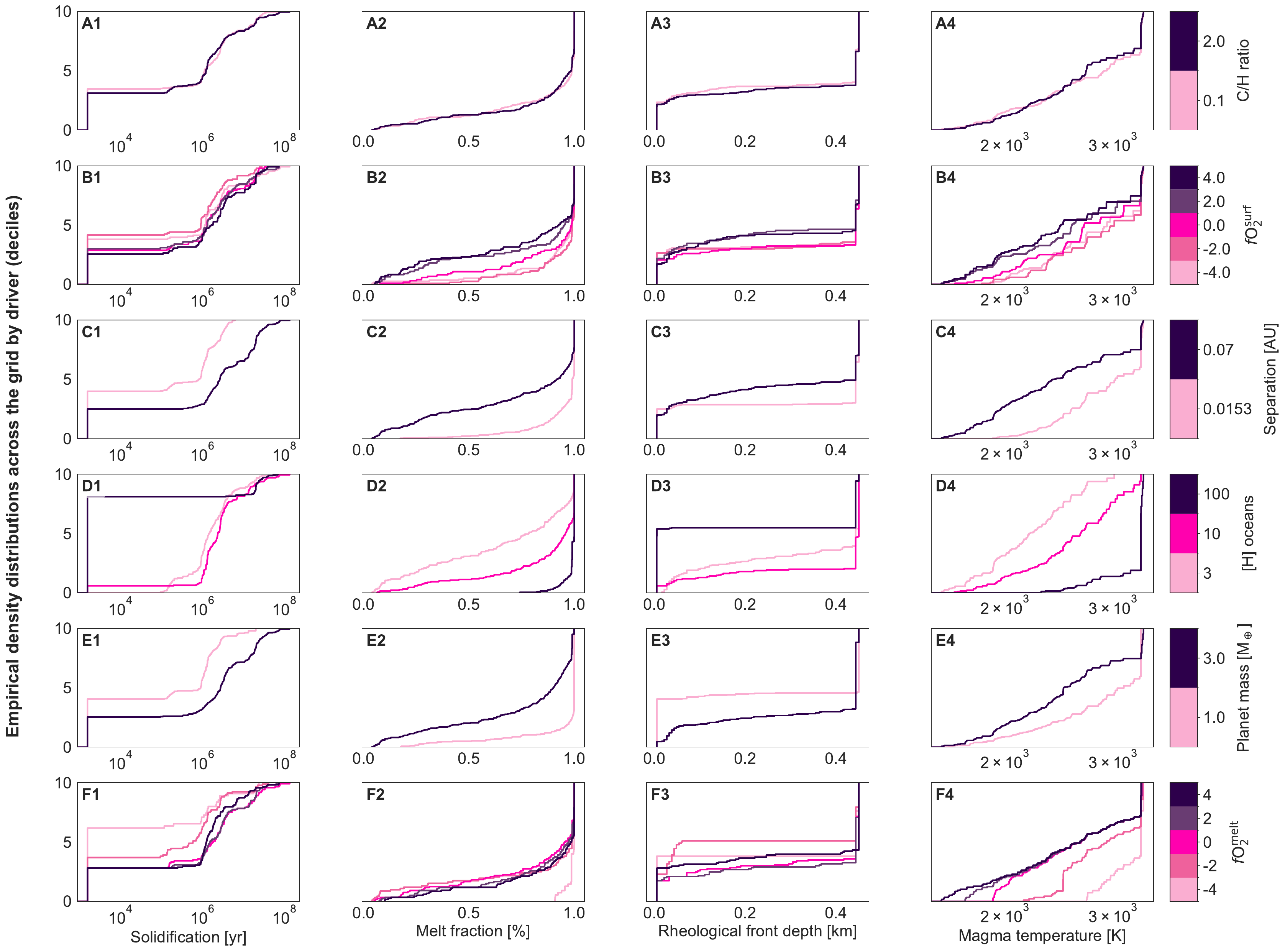}
    \caption{
    Empirical cumulative distribution function (ECDF) ``impactogram'' showing how 
    each interior driver parameter shapes the distribution of key magma ocean 
    outcomes across the full \textsc{proteus} grid. 
    Rows correspond to the six driver parameters explored in the grid: 
    C/H ratio (row~A), outgassing redox $\log f\mathrm{O}_{2}$ 
    (row~B), orbital separation (row~C), bulk water budget expressed as 
    H-oceans (row~D), planet mass (row~E), and melting-curve 
    $\Delta\mathrm{IW}$ shift (row~F). 
    Columns show the distribution of four interior outcomes: 
    solidification time (col.~1), melt fraction at the end of 
    the evolution (col.~2), rheological front depth (col.~3), and surface 
    temperature (col.~4).For each driver, all simulations at the discrete grid levels are grouped 
    and shown as ECDF curves. Colour intensity encodes the discrete values of each driver 
    (colorbars at the right of each row), with darker colours representing 
    higher grid values. Y-axis values (0-10) correspond to deciles of the 
    cumulative distribution. This figure highlights how physical and chemical parameters imprint 
    characteristic statistical signatures on magma ocean evolution. 
    Together, these ECDF provide a compact summary of the sensitivity of 
    magma ocean interior states to the multi-parameter driver space explored 
    in this study.
    }
    \label{fig:grid_impactogram}
\end{figure}

\begin{figure}[htb]
    \centering

    \includegraphics[width=0.95\textwidth]{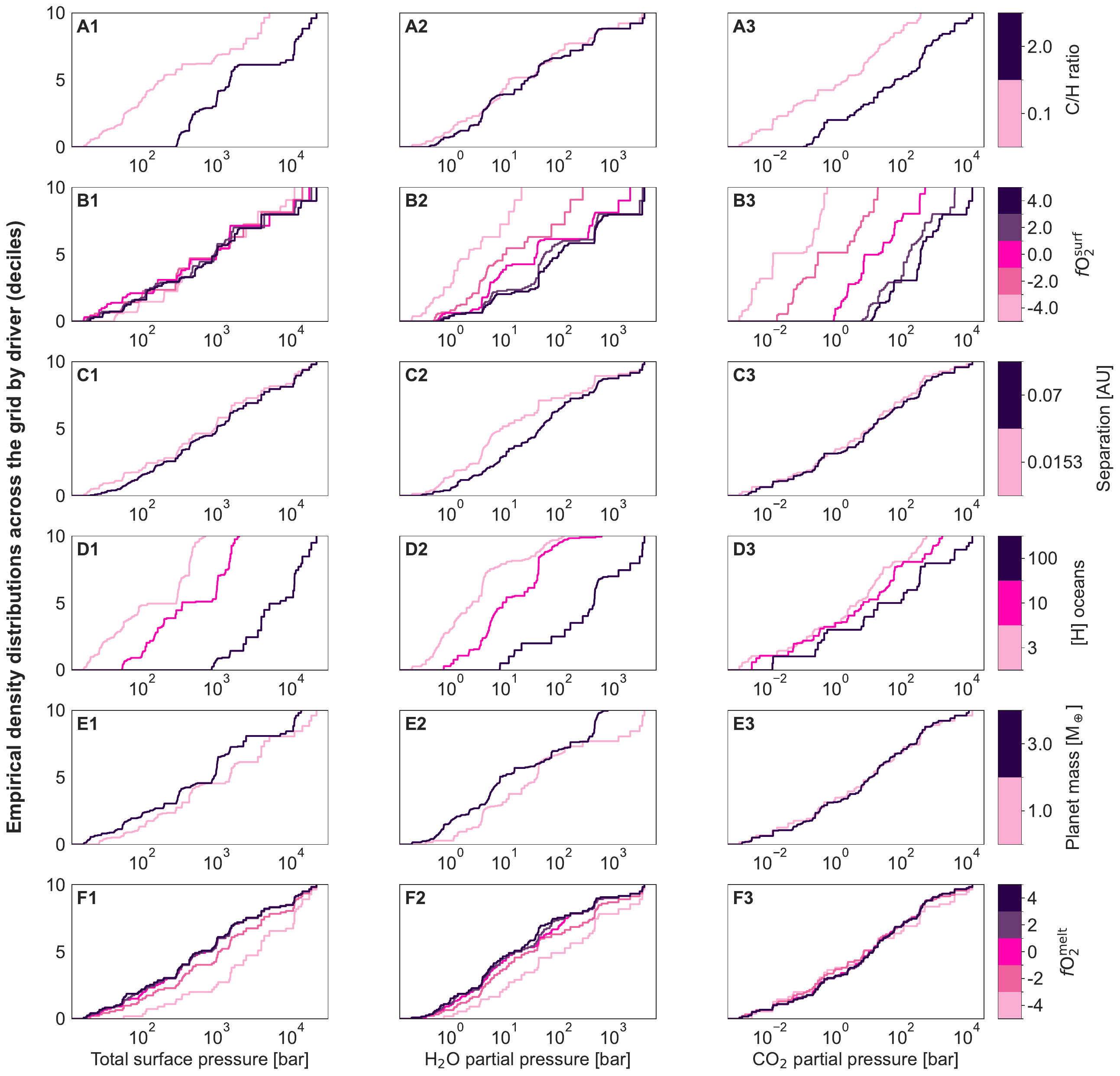}
    \caption{Same as Fig \ref{fig:grid_impactogram} but illustrating how each interior driver parameter influences the distribution of atmospheric outcomes across the full PROTEUS grid. Columns show the resulting distributions for total Surface pressure(col 1), H$_{2}$O partial pressure (col 2), and CO$_2$ partial pressure (col 3).}
    \label{cumulative_atmo}
\end{figure}

\end{document}